\documentclass[12pt]{article}
\usepackage{amsmath,amssymb,amsfonts,graphicx,cite,color,feynarts,soul}
\usepackage{epsfig}
\usepackage[normalem]{ulem}
\input{paperdef}

\graphicspath{{figs/}}

\evensidemargin \oddsidemargin
\allowdisplaybreaks

\newcommand{\SARAH}{{\tt SARAH}}

\newcommand{\FeynArts}{{\tt FeynArts}}

\newcommand{\FeynHiggs}{{\tt FeynHiggs}}
\newcommand{\NMSSMTools}{{\tt NMSSMTools}}
\newcommand{\HiggsBounds}{{\tt HiggsBounds}}
\newcommand{\GeV}{{\;\mathrm{GeV}}}
\newcommand{\mue}{{\mu_{\mathrm{eff}}}}

\newcommand{\lhcee}{LHC$_{2011}$}

\begin{document}
\thispagestyle{empty}

\def\thefootnote{\fnsymbol{footnote}}

\begin{flushright}
DESY 12--114 
\end{flushright}

\vspace{0.5cm}

\begin{center}

{\large\sc {\bf Confronting the MSSM and the NMSSM with the\\[.5em]
Discovery of a Signal in the two Photon Channel at the LHC}}

\vspace{1cm}

{\sc
R.~Benbrik$^{1}$%
\footnote{email: benbrik@ifca.unican.es}%
, M.~Gomez Bock$^{2}$%
\footnote{email: melina@ifca.unican.es}%
, S.~Heinemeyer$^{1}$%
\footnote{email: Sven.Heinemeyer@cern.ch}%
,\\[.3em] O.~St{\aa}l$^{3}$%
\footnote{email: oscar.stal@desy.de}%
, G.~Weiglein$^{3}$%
\footnote{email: Georg.Weiglein@desy.de}%
~and L.~Zeune$^{3}$%
\footnote{email: lisa.zeune@desy.de}
}

\vspace*{.7cm}

{\sl
$^1$Instituto de F\'isica de Cantabria (CSIC-UC), Santander,  Spain

\vspace*{0.1cm}
$^2$Facultad de Ciencias F\'isico-Matem\'aticas Benem\'erita Universidad
Aut\'onoma de Puebla, Pue.\ M\'exico

\vspace*{0.1cm}

$^3$DESY, Notkestra\ss e 85, D--22607 Hamburg, Germany
}

\end{center}

\vspace*{0.1cm}

\begin{abstract}
\noindent
\htg{\htb{We confront the discovery of a boson decaying to two
photons}, as reported recently by ATLAS and CMS, with the corresponding
predictions in the Minimal Supersymmetric 
Standard Model (MSSM) and the Next-to-Minimal Supersymmetric Standard
Model (NMSSM). 
We perform a scan over the relevant regions of parameter space in both
models and evaluate the MSSM and NMSSM predictions for the dominant Higgs
production channel and the photon photon decay channel.
Taking into account the experimental constraints from previous direct
searches, flavour physics, electroweak 
measurements as well as theoretical considerations, 
we find that a Higgs signal in the two photon channel with
a rate equal to, or above, the SM prediction is viable over the full
mass range $123\gev \lsim M_H \lsim 127\gev$, both in the MSSM and the
NMSSM. We find that besides the interpretation of 
a possible signal at about $125\gev$ in terms of the
lightest $\cp$-even Higgs boson, both the MSSM and the NMSSM permit also
a viable interpretation where an observed state at about $125\gev$ would
correspond to the second-lightest $\cp$-even Higgs boson in the
spectrum, which would be accompanied by another light Higgs with
suppressed couplings to $W$ and $Z$ bosons.}
\htr{We find that a significant enhancement of the $\ga\ga$ rate,
compatible with the signal strenghts observed by ATLAS and CMS, 
is possible in both the MSSM and the NMSSM, and we analyse in detail
different mechanisms in the two models that can give rise to such an
enhancement.}
We briefly discuss also our
predictions in the two models for the production and subsequent decay 
into two photons of a $\cp$-odd Higgs boson.

\end{abstract}

\def\thefootnote{\arabic{footnote}}
\setcounter{page}{0}
\setcounter{footnote}{0}

\newpage


\section{Introduction}
Two of the most important goals of the experiments at the Large Hadron
Collider (LHC) are to identify the origin of electroweak symmetry
breaking, and to search for physics effects beyond the Standard Model
(SM). The Higgs searches currently ongoing at the LHC (and previously 
carried out at the Tevatron)
address both those goals. The spectacular 
discovery of a Higgs-like particle 
with a mass around $M_H\simeq 125\gev$, which has just been announced by ATLAS
and CMS~\cite{discovery}, marks a milestone of an effort that has been 
ongoing for almost half a century and opens a new era of particle physics. 
\htg{Both ATLAS and CMS reported a clear excess around $\sim 125 \gev$
in the two photon channel as well as in the $ZZ^{(*)}$ channel, whereas
the analyses in other channels} \htr{have a lower mass resolution 
and at present are less mature.} The combined sensitivity
in each of the experiments reaches about $\sim 5 \si$. \htr{The
observed rate in the $\ga\ga$ channel turns out to be considerably above
the expectation for a SM Higgs both for ATLAS and CMS. While the
statistical significance of this 
possible deviation from the SM prediction is not sufficient at present
to draw a definite conclusion, if confirmed in the future it could be a
first indication of a non-SM nature of the new state.}
The recent announcement follows the 
tantalising hints for an excess around $M_H\simeq 125\gev$ that had
already been reported both by ATLAS~\cite{atlashiggs} and
CMS~\cite{cmshiggs,CMSUpdate}. The results of the SM Higgs searches at
the Tevatron, based on the full dataset collected by CDF and D0 have
also just been announced~\cite{SMHiggsTevFinal}, showing a broad excess
in the region around $\MH \sim 125 \gev$ that reaches a significance of
nearly $3\,\si$ and would be compatible with a signal at about 
$\MH \sim 125 \gev$.

The prime task now is clearly to study the properties of the discovered
new particle and in particular to test whether the new particle is
compatible with the Higgs boson of the SM or whether there are
significant deviations from the SM predictions, which would point
towards physics beyond the SM. The fact that the observed signal in the 
$H \to \ga\ga$ channel appears to be somewhat stronger than expected in
the SM could be a first hint in this direction, although it is
statistically not very significant up to now. This result nevertheless
serves as a strong motivation for investigating possible alternatives to
the SM where a possible signal in the $\ga\ga$ channel could be enhanced
compared to the SM case.

One of the leading candidates for
physics beyond the SM is supersymmetry (SUSY), which doubles the
particle degrees of freedom by predicting two scalar partners for all SM
fermions, as well as fermionic partners to all bosons. The most widely
studied SUSY framework is the Minimal Supersymmetric Standard Model
(MSSM)~\cite{mssm}, which keeps the number of new
fields and couplings to a minimum. The Higgs sector in particular
contains two Higgs doublets, which in the $\cp$ conserving case leads to
a physical spectrum consisting of two 
$\cp$-even, one $\cp$-odd and two charged Higgs bosons.  

Going beyond the MSSM, this model has a simple extension in the
Next-to-Minimal Supersymmetric Standard Model (NMSSM),
see e.g.~\cite{nmssmrev} for reviews. A particularly
appealing motivation for
considering the NMSSM is that it provides a solution for 
naturally associating an adequate scale to
the $\mu$ parameter appearing in the MSSM superpotential
\cite{Ellis:1989,Miller:2003ay}.
In the NMSSM, the introduction of a new singlet superfield that only
couples to the 
Higgs sector gives rise to an effective $\mu$-term, generated in a
similar way  as the Yukawa mass terms of fermions through its
vacuum expectation value. The new field must be a gauge singlet, since 
the parameter
$\mu$ carries no 
$SU(3)_{\rm C}\times SU(2)_{\rm I}\times U(1)_{\rm Y}$
quantum numbers.
This effective $\mu$-term is linked
dynamically to the electroweak scale. 
The additional degrees of freedom from the singlet add
to the NMSSM particle spectrum. In the case where
$\cp$ is conserved, which we assume throughout the paper, the states
in the Higgs sector 
can now be classified as three $\cp$-even Higgs bosons, $\Hi$ ($i = 1,2,3$),
two $\cp$-odd Higgs bosons, $\Aj$ ($j = 1,2$), and the charged Higgs
boson pair $H^\pm$.  In addition, the SUSY partner of the singlet Higgs
(called the singlino) extends the neutralino sector (to a total of five
neutralinos). 

The extended parameter space of the NMSSM as compared to the MSSM
case gives rise to a rich and interesting phenomenology, in particular
related to states in the Higgs sector that can be relatively light.
As an example, Higgs-to-Higgs decays can be
open (as in 
some of the benchmarks presented in \cite{Djouadi:2008uw}), with
potential consequences for Higgs searches at the LHC. This
non-trivial phenomenology of the NMSSM led in fact to the situation that 
it has not been possible to establish a general ``no-loose'' theorem
for NMSSM Higgs discovery at the LHC~\cite{Noloose} (similar conclusions
were obtained for the case of the MSSM with complex parameters, 
see e.g.~\cite{MSSMcomplex}.) Scenarios with a light $\cp$-even Higgs
boson with SM-like couplings to gauge bosons that however decays in an
unusual way into a pair of very light $\cp$-odd
Higgs bosons $\Ae$ ($\mAe < 2 m_b$), sometimes called 
``ideal Higgs scenarios''~\cite{DermisekGunion}, have found particular
attention. The case of a light $\cp$-even Higgs in the NMSSM with suppressed
couplings to gauge bosons has been discussed in~\cite{Stal:2011cz}.
A summary of 
phenomenological results for the NMSSM Higgs sector can be found 
in~\cite{Ellwanger:2011sk}. The presence of an additional neutralino
can have implications for dark matter, which may in this case
contain a sizeable
singlet fraction (see \cite{nmssmrev} and references therein) or---like
the new Higgs bosons---be very light \cite{LightNeutralino}.

Inspired by the spectacular discovery of a new particle in the LHC data
that manifests itself in particular as a signal of at least (but
possibly higher than) SM strength in the $\gamma\gamma$ channel, we 
investigate the corresponding predictions in both the MSSM and the
NMSSM, and compare them to the SM case. In particular, we evaluate the
predictions for 
the production of a MSSM or NMSSM Higgs boson via gluon fusion,
the main production channel at the LHC,
followed by the decay into two photons, the channel with the largest
significance in the results that have just been reported. As a main
focus, we analyse
potential enhancements of the production cross section times branching
ratio over the corresponding SM prediction and we confront those
predictions with the experimental data. We discuss in detail how 
an enhanced $\ga\ga$ rate can be realised in
the MSSM, and which {\em additional} mechanisms for an enhancement can
occur in the NMSSM. We find that sizable enhancements of the $\ga\ga$
rate are possible in both models, accompanied by or without a suppression
of the $WW^{(*)}$ decay mode. It is interesting to note that in both
models the signal can be interpreted \emph{either} as the lightest 
$\cp$-even Higgs
\emph{or} as the second-lightest $\cp$-even Higgs. The latter
interpretation would imply that there would be an additional Higgs boson
present 
with a mass below $125\gev$ and suppressed couplings to gauge bosons.

\medskip
This paper is organised as follows: in \refse{sect:nmssm}
we give a short introduction
to the NMSSM and specify our notation. In \refse{sect:loop} we describe the
more technical issues of the loop calculations involved: the
framework for one-loop calculations in the NMSSM, the tools used for
the numerical evaluation of observables, and we discuss 
constraints on the NMSSM
parameter space. The main results are given in Section~\ref{sect:results}.
We show the numerical results on loop-induced Higgs production and
decays in the MSSM and NMSSM, which are compared to 
corresponding results in the SM (where relevant). Finally,
the model predictions are confronted with the most recent results from
the LHC Higgs searches, namely the discovery of a Higgs-like
particle, and the implications are discussed. 
Section~\ref{sect:conclusions}
contains our conclusions.  


\section{The NMSSM}
\label{sect:nmssm}

\subsection{General considerations}
Like the MSSM, the NMSSM contains two Higgs doublet superfields
$\hat{H}_1$, $\hat{H}_2$, but in addition there is also a singlet scalar
superfield $\hat{S}$.%
\footnote{Quantities with a hat denote superfields; fields without hats
  are their corresponding scalar components.}%
~If one only considers the new interaction between the three scalar
fields which gives rise to the effective $\mu$-term a Peccei-Quinn (PQ)
symmetry is introduced. This may be avoided by considering a term cubic
in the new singlet superfield, with a coupling strength $\kappa$.  
This cubic term explicitly breaks the PQ symmetry, providing a
mass to the scalar $\cp$-odd (would be) axion. In order to reduce the
interactions permitted in a general superpotential 
we assume a discrete $Z_3$ symmetry.
In this so-called $Z_3$-symmetric version of the NMSSM the
superpotential is assumed to be scale invariant (preventing terms linear
or quadratic in $\hat{S}$), 
 thus it takes the general form
\begin{equation}
W^{\mathrm{NMSSM}}=y_D\hat{Q}\cdot \hat{H}_1 \hat{D}^c+y_L\hat{L}\cdot
\hat{H}_1 
\hat{E}^c+y_U\hat{Q}\cdot \hat{H}_2 \hat{U}^c+\lambda \hat{S} \hat{H}_2\cdot
\hat{H}_1 + \frac{1}{3}\kappa\hat{S}^3, 
\label{eq:superpot}
\end{equation}
where contraction of $SU(2)$ indices is denoted by a dot product,
i.e.~$\hat{Q}\cdot \hat{H}\equiv\epsilon_{ij}\hat{Q}^i\hat{H}^j$ with
$\epsilon_{ij}=-\epsilon_{ji}$ and $\epsilon_{12}=1$. 
Assuming supersymmetry breaking by an unspecified mechanism, 
there are new
soft breaking terms compared to the MSSM that need to be considered,
\begin{equation}
V_{\mathrm{soft}}^{\mathrm{NMSSM}}=
V^{\mathrm{MSSM}}_{\mathrm{soft}}+m_S^2|S|^2+\lambda 
A_\lambda SH_2\cdot H_1+\frac{1}{3}\kappa A_\kappa S^3+\mathrm{h.c.} .
\label{eq:Vsoft}
\end{equation}
Ignoring the possibility of $\cp$ violation, as we do throughout this
work, all parameters can be assumed real.


\subsection{The Higgs sector}

The explicit form of the Higgs potential is obtained from the
Higgs $F$-terms, $D$-terms and the soft SUSY-breaking terms as  
$V_{H}=V_{F}+V_{D}+V_{\mathrm{soft}}$.
The $F$-terms arise from the first derivative of the
$Z_{3}$-invariant superpotential given in
Eq.~\eqref{eq:superpot}. As the singlet $S$ is assumed to only couple to
the Higgs, there is no coupling to the gauge bosons, and the $D$-terms
are the same as in the MSSM. The soft breaking terms relevant to the Higgs
sector are given in Eq.~\eqref{eq:Vsoft}. Combining everything, we get
the scalar Higgs potential of the NMSSM as%
\footnote{We use a notation that closely follows the one specified
in \cite{mhcMSSMlong} for the MSSM.}%
\begin{align}
  \VHiggs &= 
  m_1^2 |H_{1}|^2 
+ m_2^2 |H_{2}|^2 
+ m_S^2 |S|^2 + |\la  \eps_{ij} (H_{2i}H_{1j}) + \ka S^2|^2
+ |\la|^2 |S|^2 \left(|H_{1}|^2+|H_{2}|^2\right) \non \\
&\quad + \KL \la \Ala \eps_{ij} (H_{2i}H_{1j}) S
            +\ed{3} \ka \Aka S^3 + {\rm c.c.} \KR +\frac{1}{8}(g_1^2+g_2^2)\left(|H_{1}|^2-|H_{2}|^2\right)^2 
    + \frac{1}{2} g_2^2 | H_{1i}^* H_{2i} |^2. 
    \label{eq:higgspot}
\end{align}
\noindent
The indices $\{i,j\}=\{1,2\}$ refer to the respective Higgs doublet
component (summation over $i$ and $j$ is understood). 
The $\cp$-invariant Higgs potential $\VHiggs$ contains the real soft
breaking parameters 
$m_1^2$, $m_2^2$, $m_S^2$, 
 $\Ala$, $\Aka$, as well as the superpotential trilinear couplings $\la$
and $\ka$, and the $U(1)_{\rm Y}$ and $SU(2)_{\rm I}$
coupling constants $g_1$ and $g_2$.

After spontaneous symmetry breaking the two Higgs doublets can be
expanded similarly to the MSSM. In components they are given by 
\begin{equation}
H_1 = \left(\begin{array}{c}
v_1+\frac{1}{\sqrt{2}}\left(\phi_1-\mathrm{i}\chi_1\right)\\
-\phi_1^-
\end{array}\right),
\qquad
H_2 = \left(\begin{array}{c}
\phi_2^+\\
v_2+\frac{1}{\sqrt{2}}\left(\phi_2+\mathrm{i}\chi_2\right)
\end{array}\right),
\end{equation}
and the ratio of the two vacuum expectation values defines
\begin{equation}
\tb\equiv\frac{v_2}{v_1}.
\end{equation}
The complex scalar singlet can likewise be expanded as
\begin{equation}
S=v_s+\frac{1}{\sqrt{2}}\left(\phi_s+\mathrm{i}\chi_s\right).
\end{equation} 
According to \refeq{eq:superpot} a non-zero vacuum expectation value 
$v_s$ for the singlet gives rise to
an effective $\mu$ parameter (cf.~the bilinear
term $W_2\sim \mu \hat{H}_2\cdot \hat{H}_1$ in the MSSM superpotential) 
\begin{equation}
\mueff= \la v_s~.
\label{def-mueff}
\end{equation}
Since the scale for $\mueff$ is set by $v_s$, a value at the
electroweak scale is natural.
Minimising the Higgs potential, the mass parameters $m_1^2$ and $m_2^2$
appearing in \refeq{eq:higgspot} are replaced by the vacuum expectation
values $v_1$ and $v_2$ (or, equivalently, $v=\sqrt{v_1^2+v_2^2}\simeq
174 \gev$ and $\tb$). Similarly, the singlet soft mass
$m_S^2$ can be expressed in terms of $v_s$. 

The bilinear part of the Higgs potential is then given by
\begin{align}
\VHiggs &= \tedz \begin{pmatrix} \phi_1,\phi_2, \phi_S \end{pmatrix} 
\matr{M}_{\phi\phi\phi}
\begin{pmatrix} \phi_1 \\ \phi_2 \\ \phi_S \end{pmatrix} +
\tedz \begin{pmatrix} \chi_1, \chi_2, \chi_S \end{pmatrix}
\matr{M}_{\chi\chi\chi}
\begin{pmatrix} \chi_1 \\ \chi_2 \\ \chi_S \end{pmatrix} +
\begin{pmatrix} \phi^-_1,\phi^-_2  \end{pmatrix}
\matr{M}_{\phi^\pm\phi^\pm}
\begin{pmatrix} \phi^+_1 \\ \phi^+_2  \end{pmatrix} + \cdots, 
\end{align}
with the mass matrices
$\matr{M}_{\phi\phi\phi}$, $\matr{M}_{\chi\chi\chi}$ 
and $\matr{M}_{\phi^\pm\phi^\pm}$ for the three $\cp$-even neutral scalars
$(\phi_1,\phi_2,\phi_s)$, three $\cp$-odd neutral scalars
$(\chi_1,\chi_2,\chi_s)$, and two charged pairs
$(\phi_1^\pm,\phi_2^\pm)$, respectively.
At tree-level, the elements of the mass matrix
$\matr{M}_{\phi\phi\phi}$ for the $\cp$-even Higgs bosons is given in the basis
$(\phi_1,\phi_2,\phi_s)$ by 
\begin{equation}
\begin{aligned}
\left(\matr{M}_{\phi\phi\phi}\right)_{11} &=
M_Z^2\cos^2\beta+B\mueff\tb , \\
\left(\matr{M}_{\phi\phi\phi}\right)_{22} &=
M_Z^2\sin^2\beta+B\mueff\cot\beta , \\
\left(\matr{M}_{\phi\phi\phi}\right)_{33} &= \frac{\la^2 v^2 \Ala}{\mueff}
\cos\beta\sin\beta+K\mueff\left(\Aka+4K\mueff\right) , \\
\left(\matr{M}_{\phi\phi\phi}\right)_{12} &= \left(2\la^2
v^2-M_Z^2\right)\cos\beta\sin\beta-B\mueff , \\
\left(\matr{M}_{\phi\phi\phi}\right)_{13} &= \la
v\left[2\mueff\cos\beta-\left(B+K\mueff\right)\sin\beta\right] , \\
\left(\matr{M}_{\phi\phi\phi}\right)_{23} &= \la
v\left[2\mueff\sin\beta-\left(B+K\mueff\right)\cos\beta\right] ,
\end{aligned}
\end{equation}
where we have introduced $K\equiv \ka/\la$ and $B\equiv
\Ala+K\mueff$, and $\MZ$ denotes the mass of the $Z$~boson.

The $\cp$-odd mass matrix is simplified significantly by introducing an
effective doublet mass 
\begin{equation}
\map^2=\frac{B\mueff}{\Sbe\Cb},
\label{eq:MA2}
\end{equation} 
which corresponds to the mass of the single $\cp$-odd Higgs boson in the MSSM
limit. In the basis ($\chi_1,\chi_2,\chi_s$) this gives 
\begin{equation}
\begin{aligned}
\left(\matr{M}_{\chi\chi\chi}\right)_{11} &=\map^2\sin^2\beta , \\
\left(\matr{M}_{\chi\chi\chi}\right)_{22} &=\map^2\cos^2\beta , \\
\left(\matr{M}_{\chi\chi\chi}\right)_{33} &=\frac{\la^2
v^2}{\mueff}\left(B+3K\mueff\right)\cos\beta\sin\beta-3
\Aka K\mueff , \\
\left(\matr{M}_{\chi\chi\chi}\right)_{12} &=\map^2\sin\beta\cos\beta , \\
\left(\matr{M}_{\chi\chi\chi}\right)_{13} &=\la v\left(B-3K\mueff\right)\sin\beta , \\
\left(\matr{M}_{\chi\chi\chi}\right)_{23} &=\la v\left(B-3K\mueff\right)\cos\beta 
\end{aligned}
\end{equation}
for the elements of the $\cp$-odd mass matrix. 

\newcommand{\Ueven}{U^H}
\newcommand{\Uodd}{U^A}
\newcommand{\Uchar}{U^C}
The mass eigenstates in lowest order follow from 
unitary transformations of the original fields, 
\begin{align}
\label{eq:RotateToMassES}
\begin{pmatrix} \He \\ \Hz \\ \Hd \end{pmatrix} 
= \Ueven \cdot
\begin{pmatrix} \phi_1 \\ \phi_2 \\ \phi_S \end{pmatrix}, \quad
\begin{pmatrix} \Ae \\ \Az \\ G \end{pmatrix} 
= \Uodd \cdot
\begin{pmatrix} \chi_1 \\ \chi_2 \\ \chi_S \end{pmatrix}, \quad
\begin{pmatrix} H^\pm \\ G^\pm \end{pmatrix} = \Uchar \cdot
\begin{pmatrix} \phi_1^\pm \\ \phi_2^\pm \end{pmatrix} .
\end{align}

The matrices $\Ueven$, $\Uodd$ and
$\Uchar$  
transform the neutral $\cp$-even, $\cp$-odd and charged Higgs fields,
respectively, such that the resulting mass matrices 
\begin{equation}
\matr{M}_{hhh}^{\rm diag} = \Ueven
\matr{M}_{\phi\phi\phi} {\Ueven}^{\dagger}, \;
\matr{M}_{aaG}^{\rm diag} = \Uodd
\matr{M}_{\chi\chi\chi} {\Uodd}^{\dagger} 
\; {\rm and} \;
\matr{M}_{H^\pm G^\pm}^{\rm diag} = 
\Uchar \matr{M}_{\phi^\pm\phi^\pm}
{\Uchar}^{\dagger} 
\label{eq:RotMat}
\end{equation}
are diagonal in the basis of the transformed fields.
The new fields correspond to the three neutral $\cp$-even Higgs bosons
$\He$, $\Hz$ and $\Hd$ with $\mHe \le \mHz \le \mHd$, 
the two $\cp$-odd Higgs bosons $\Ae$ and $\Az$ with $\mAe \le \mAz$, 
the charged pair $H^\pm$ and the Goldstone bosons $G$ and
$G^\pm$. 
The tree-level mass of the physical charged Higgs boson is related to
the $\cp$-odd doublet mass through (with $\MW$ denoting the mass of
the $W$~boson)
\begin{equation}
\MHp^2=\map^2+M_W^2-\la^2 v^2,
\label{eq:MHp2}
\end{equation}
which shows that $\MHp$ may be lower than its corresponding MSSM value.


\subsection{The neutralino and chargino sectors}
In addition to the extended scalar sector, the supersymmetric partner of
the singlet Higgs field---the singlino $\tilde{S}$---leads to the
presence of fifth neutralino in the NMSSM spectrum. As a result, the
neutralino mass matrix in the basis $(\tilde{B}, \tilde{W}^0,
\tilde{H}_1^0, \tilde{H}_2^0, \tilde{S})$ is given at tree-level by 
\begin{equation}
\matr{M}_{\neu{}}=\left(\begin{array}{ccccc}
M_1 & 0 & \displaystyle -g_1\frac{v_1}{\sqrt{2}}& \displaystyle
g_1\frac{v_2}{\sqrt{2}}& 0 \\ 
0 & M_2 & \displaystyle  g_2\frac{v_1}{\sqrt{2}}& \displaystyle
-g_2\frac{v_2}{\sqrt{2}}& 0 \\ 
\displaystyle -g_1\frac{v_1}{\sqrt{2}} & \displaystyle
g_2\frac{v_1}{\sqrt{2}} & 0 & -\mue& -\lambda v_2\\ 
\displaystyle  g_1\frac{v_2}{\sqrt{2}} & \displaystyle
-g_2\frac{v_2}{\sqrt{2}} & -\mue & 0& -\lambda v_1 \\ 
0 & 0 & -\lambda v_2 & -\lambda v_1 & 2K\mue
\end{array}\right).
\end{equation}
It can be diagonalised by a single unitary (complex) matrix $N$ such that
\begin{equation}
\mathrm{diag}(\mneu{1}, \mneu{2},\ldots)=N^* \matr{M}_{\neu{}} N^\dagger
\end{equation}
is real, positive definite, and with the mass eigenvalues ordered as 
$\mneu{i} \le \mneu{j}$ for $i<j$.

Since no new charged degrees of freedom are introduced, the chargino
sector of the NMSSM is identical to that in the MSSM. 


\subsection{Squarks}
The squark sector of the NMSSM is unchanged with respect to the MSSM. 
Assuming minimal flavour violation and $\cp$ conservation, the mass
matrix of the two squarks of the same flavour, $\sql$ and $\sqr$, 
is given by 
\begin{align}
\matr{M}_{\sq} =
\begin{pmatrix}
        \MsqL^2 + \mq^2 + \MZ^2 \CZb (I_3^q - Q_q \sw^2) & \mq \; \Xq \\
        \mq \; \Xq    & \MsqR^2 + \mq^2 + \MZ^2 \CZb Q_q \sw^2
\end{pmatrix} ,
\label{squarkmassmatrix}
\end{align}
with
\begin{align}
\Xq &= A_q - \mueff \{\CTb, \tb\} ,
\label{squarksoftSUSYbreaking}
\end{align}
where $\{\CTb, \tb\}$ applies for up- and down-type squarks, respectively.
In these equations,  $\MsqL$, $\MsqR$ are (real) soft SUSY-breaking
masses (as a consequence of $SU(2)$ symmetry, $\MsqL$ is equal for
the members of an $SU(2)$ doublet; in the numerical analysis below we assume 
a universal value
$\msusy= \MsqL=\MsqR$), while $A_q$ is the soft SUSY-breaking trilinear
coupling and $\mueff$ the effective $\mu$ parameter as defined above.
$I_3^q$ and $Q_q$ denote the weak isospin and the electric charge
  of the quark, respectively, and $\sw = \sqrt{1 - \cw^2}$ with 
$\cw = \MW/\MZ$.
The squark mass eigenstates are obtained by the unitary transformation
\BE
\VL \sqe \\ \sqz \VR = \matr{U}_{\sq} \VL \sql \\ \sqr \VR
\end{equation}
with
\BE
U_{\sq} = \ML \costq & \sintq \\ -\sintq & \costq \MR
               ,\quad
U_{\sq} U_{\sq}^{\dagger} = \unity~,
\end{equation}
and the mass eigenvalues are given by
\begin{equation}
\begin{aligned}
m_{\tilde q_{1,2}}^2 = \mq^2
  + \edz 
&\Biggl[
\MsqL^2 + \MsqR^2 + I_3^q \MZ^2 \CZb \\&
           \mp \sqrt{[\MsqL^2 - \MsqR^2 + \MZ^2 \CZb(I_3^q -2 Q_q
  \sw^2)]^2 + 4 \mq^2 \Xq^2}~\Biggr]~.
\end{aligned}
\end{equation}



\section{Higher-order corrections in the (N)MSSM}
\label{sect:loop}

\subsection{General considerations}

The investigations in the present paper are part of a bigger
activity that we have undertaken in order to provide precise theoretical
predictions for relevant observables in the NMSSM that can be used for
comparing the NMSSM phenomenology with the SM, the MSSM and other 
scenarios of physics beyond the SM and for confronting the predictions
with the available data. 
In order to investigate to what extent the Higgs search results can
discriminate between different scenarios of physics at the Terascale,
precise theoretical
predictions both within the SM and possible alternatives of it are
needed. In particular, if small deviations from the SM predictions are
probed it is crucial to treat the considered model of new physics at the
same level of precision to enable an accurate analysis and comparison. 
In the MSSM Higgs sector higher-order contributions are known to give 
numerically large effects (see, e.g.,
\cite{MSSMHiggsRev,MSSMHiggsRev2}), and the same also holds for the
NMSSM (see, e.g., \cite{NMSSMHiggs}).
For many observables it is therefore 
necessary to include corrections beyond leading order in the
perturbative expansion to obtain reliable results. 
It is planned that
the results presented in this paper together with other results in the 
NMSSM Higgs sector~\cite{WIP} will be
implemented into a new version of the Fortran code
{\tt FeynHiggs}~\cite{feynhiggs,mhiggslong,mhiggsAEC,mhcMSSMlong}, which
so far is restricted to predictions for Higgs physics in the SM and the 
MSSM.
A few public codes already exist for numerical NMSSM calculations. The
by far most widely used is \NMSSMTools 
\cite{Ellwanger:2004xm}, which consists of subpackages to calculate the
NMSSM spectrum, constraints on the parameter space, as well as Higgs
decays and decay modes of sparticles
\cite{Das:2011dg}. Recently also an extension applicable to the NMSSM
of the program {\tt SPheno}~\cite{SPheno} became available, which makes
use of model implementations 
generated with \SARAH~\cite{SARAH}.

Since the NMSSM extends the MSSM in the Higgs and the 
neutralino sectors, differences to the MSSM are best
probed in these two sectors. 
The processes playing the main role in the reported discovery at
the LHC, production via gluon fusion and decay into two photons, 
are in fact processes that are particularly sensitive to possible 
deviations between the SM, the MSSM and the NMSSM. 
The one-loop predictions for those processes 
correspond to the leading-order contributions, which are IR- and also
UV-finite without renormalisation (for a recent discussion of the 
renormalisation of the NMSSM Higgs sector, see \cite{Ender:2011qh}), so
that the set-up mentioned above can immediately be applied for the
investigation of the processes that are more important for NMSSM Higgs
phenomenology at the LHC. 
After the initial announcements of an excess in the LHC Higgs searches
by ATLAS~\cite{atlashiggs} and CMS~\cite{cmshiggs} there was already a
considerable activity in the literature concerning possible 
interpretations of the results in various versions of the MSSM
\cite{Mh125our,Mh125gaga,Mh125}, and also in the NMSSM
\cite{Ellwanger2,Mh125NMSSM,Cao:2012fz}. We comment on specific 
differences and similarities between those results and ours below.

\medskip
The calculation of loop diagrams,
often involving a large number of fields, is a tedious and
error-prone task if done by hand. This is true in particular for
theories beyond the SM where the number of fields is significantly
increased. For one-loop calculations, as will be the focus in the
following, computer methods with a high degree of automation have
been devised to simplify the work. However, most of the available tools
so far have focused on calculations either in the SM or the MSSM.
In order to facilitate loop calculations in the NMSSM, it is useful
to employ the well-established public tools
{\tt FeynArts}~\cite{FeynArts}, 
{\tt FormCalc}~\cite{Hahn:1998yk,FormCalc23} and
{\tt LoopTools}~\cite{Hahn:1998yk,LoopToolsOther}, with which one-loop
calculations can 
be carried out with a high degree of automation. 
The program {\tt FeynArts}~\cite{FeynArts} can be used to generate and
draw the Feynman diagrams to a given order for the
process under study, based on the information about the particle content
and interactions that is supplied in a so-called model file. From the 
Feynman rules a mathematical expression for the corresponding amplitudes
is generated. For one-loop
amplitudes, the analytic simplifications, trace evaluation, tensor
decomposition, etc.\ can then
be carried out by {\tt FormCalc}~\cite{Hahn:1998yk,FormCalc23}, 
which combines the speed of 
{\tt FORM}~\cite{form} with the more user-friendly interface of
Mathematica. If the necessary information about the particles involved
in the considered process (mass eigenvalues, mixing angles, etc.) is
provided, the numerical evaluation of the {\tt FormCalc} output can be
carried out with {\tt LoopTools}~\cite{Hahn:1998yk,LoopToolsOther}. In
the standard 
distribution, {\tt FeynArts} contains model files for the SM
(including several variants), a general two-Higgs-doublet model, and an
implementation of the MSSM.

With the goal of treating the NMSSM at the same level of accuracy as the
MSSM, we are currently developing a framework that enables the
computation of one-loop processes in the NMSSM in a highly automated
way. This consists on the one hand of the input on the particle content,
the interaction vertices, etc.\ that is needed for use with the packages 
{\tt FeynArts} and {\tt LoopTools}, and on the other hand of an
appropriate renormalisation prescription for the NMSSM. In the present
paper we will focus on loop-induced processes, for which no
renormalisation is needed. A detailed discussion of our renormalisation
prescription for the NMSSM will be presented elsewhere.


\subsection{Implementation}

As a first step we have compiled a new
{\FeynArts}~model file for the NMSSM. Since it is foreseen that the same
framework will be applied for several calculations beyond those
presented in this paper, we describe our NMSSM
implementation in some detail here. The basis for the model file itself
--- defining the particle content and interactions of the NMSSM (as
described in \refse{sect:nmssm}) in a general $R_\xi$ gauge --- is
generated with the help
of the program {\tt SARAH}~\cite{SARAH}. This program
can be used to generate {\tt FeynArts} model files, as well as output for
many other programs, for any supersymmetric theory starting from its
superpotential. For consistency checks, we also use an independent NMSSM
model file generated with {\tt FeynRules}~\cite{FeynRules}. 
Starting from the output of {\tt SARAH} we have introduced 
the standard
nomenclature of {\tt FormCalc} to activate its internal MSSM
simplifications, applied unitarity relations to mixing
matrices and couplings, and implemented some further improvements. 
These modifications, besides greatly improving the speed at which 
{\tt FormCalc} performs one-loop calculations of NMSSM amplitudes,
are essential for instance for verifying the cancellation of UV
divergences at the algebraic level.

To enable the numerical evaluation of observables, such as decay widths and
cross sections, the analytic amplitudes of {\tt FormCalc} can be
exported to Fortran code. These must be supplemented with a so-called
driver program to compute the necessary quantities (masses, mixings,
etc.) from the fundamental parameters of the theory. The driver codes
also provide standard facilities for numerical integration and the
evaluation of master one-loop integrals through 
{\tt LoopTools}. 
We have developed such a driver program for the NMSSM,
which in its present state allows for Higgs and sparticle masses to be
calculated either at tree-level (following \refse{sect:nmssm}), or using 
\NMSSMTools\ (here we used version 2.3.5)~\cite{Ellwanger:2004xm} 
linked through a custom interface.  

The NMSSM driver also offers the possibility to impose restrictions on
the NMSSM parameter space resulting from the evaluation of various
experimental or theoretical constraints.
For instance, the constraints implemented in
\NMSSMTools\ can be accessed, and direct constraints on the extended
Higgs sector are available through an interface to
\HiggsBounds~\cite{Bechtle:2011sb}. More details on the
different constraints and how they are evaluated is given in 
\refse{sect:constraints} below.


\subsection{Verification}

We have performed several tests on the model file to verify
the NMSSM implementation, in particular for the Higgs sector which is
most relevant for the present work. The analytical expressions for the
Feynman rules for the interaction vertices of the NMSSM
obtained from {\tt SARAH} have been compared to the
independent {\tt FeynRules} output. They have also been compared
(analytically) in the MSSM limit to the corresponding vertices in the
default MSSM implementation distributed with {\tt FeynArts}.  

A number of tree-level processes have been analysed numerically ---
including the decays of Higgs bosons and neutralinos --- to test the
mixing properties of the singlet state
in the NMSSM. Comparing these to the results of
\NMSSMTools\ and {\tt NMSDECAY}~\cite{Das:2011dg}, we find overall good
agreement with those previously obtained results after correcting for
differences due to QCD corrections and the running of gauge couplings.  

\begin{figure}[htb!]
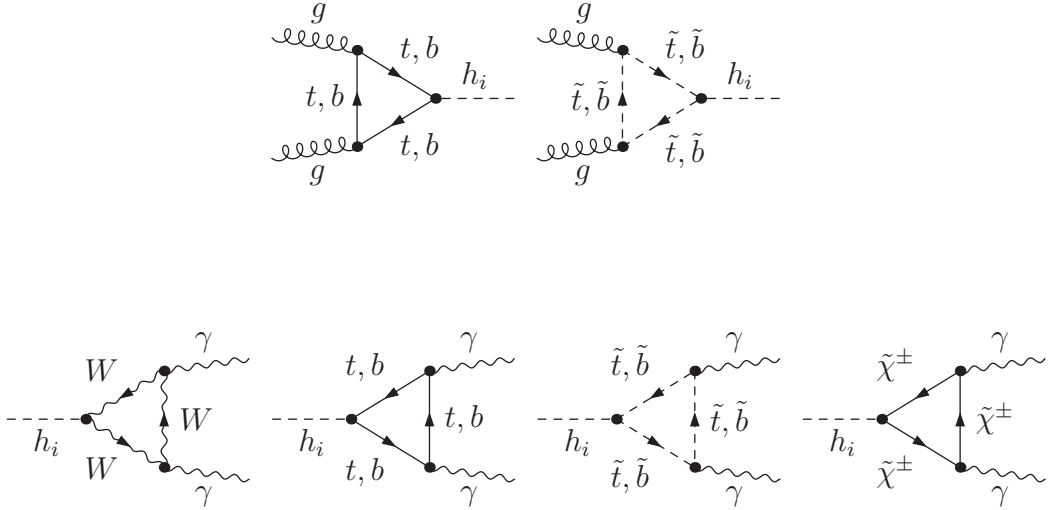

\centering
    \input{GGHiggs}
    \input{GagaHiggs}
\caption{Generic diagrams contributing to $gg \to h_i$ (upper row) and
to $h_i \to \ga\ga$ (lower row), where $h_i$ denotes any neutral
$\cp$-even Higgs boson in the (N)MSSM.}
\label{fig:generic}
\end{figure}

A further, extensive, and non-trivial test of the working NMSSM
implementation is provided by the results for the processes that
are induced at the one-loop level in the NMSSM. 
We have evaluated \order{50} $1 \to 2$ processes
and \order{100} $2 \to 2$ processes of this type
and checked them successfully 
for their UV- and IR-finiteness. This includes the Higgs production and
decay modes that are phenomenologically most relevant at the LHC, 
which are described and analysed numerically in detail below.
As an example we show in \reffi{fig:generic} generic Feynman
diagrams contributing to $gg \to h_i$ (upper row), where $h_i$ denotes
any neutral $\cp$-even Higgs boson in the (N)MSSM, as well as to 
$h_i \to \ga\ga$ (lower row). These types of diagrams have been
evaluated and used for our numerical analysis described in
\refse{sect:results}. 

\subsection{Constraints on the parameter space}
\label{sect:constraints}

Before moving on to our numerical analysis, we briefly
discuss the various phenomenological constraints which exist on the 
parameter space of the NMSSM. 
As already mentioned, our computational framework
allows for all these constraints to be evaluated and applied in
connection with the calculation of arbitrary NMSSM observables. In this
way reliable predictions can be obtained which are in agreement with the
present experimental results. 

\subsubsection*{Theoretical constraints}

Constraints originate from the requirement of a viable physical minimum
of the Higgs potential. The physical minimum, with non-vanishing vacuum
expectation values (vevs) for the two Higgs doublets $H_1$ and $H_2$,
should be lower than the local minimum with vanishing vevs for the two
Higgs doublets. Furthermore the physical vacuum should have a non-zero
singlet vev to be able to generate the $\mueff$ parameter. In the NMSSM,
the correct pattern of a symmetry breaking absolute minimum is ensured
approximately by the condition 
$\Aka^2 \geq 9\,m_{S}^2$~\cite{Ellwanger:1996gw}. For each point
considered in the NMSSM 
parameter space, we verified numerically (using \NMSSMTools) that the Higgs
potential is bounded from below and stable. 

Another requirement is that there be no Landau pole for any of the
running couplings $\la$, $\ka$, $y_t$ and $y_b$ below
$M_{\mathrm{GUT}} \approx 2\times 10^{16} \gev$. 
The renormalisation group equations for the NMSSM are known to two-loop order~\cite{King:1995vk}. 
The constraint of perturbativity up to a very high
scale restricts the range of $\la$ and $\ka$. Values of these
parameters in the perturbative regime at the GUT scale lead
to comparably small values at the weak scale, which may be combined to
give the approximate upper bound~\cite{Miller:2003ay} 
\begin{equation}
\la^2 + \ka^2 \le 0.5.
\end{equation}
In the parameter scan below we choose the ranges for $\lambda$ and 
$\kappa$ to respect this limit.

\subsubsection*{Bounds from direct Higgs searches}
The limits from Higgs searches at LEP, the Tevatron, and the
\htg{2011 LHC data} put bounds on the Higgs masses and couplings. 
\htg{The LHC bounds have now been updated~\cite{discovery} and allow a SM-like
Higgs boson only in a small window around $\sim 125 \gev$.
As strict bounds on the parameter space we will consider the} \htr{data
from LEP and the Tevatron as well as the LHC data presented in 2011}
(referred to as \lhcee) and comment on the most recent limits
as presented in~\cite{discovery} separately. 

In the NMSSM, or any 
other theory with a Higgs sector different from that in the SM, the
lower limit on the Higgs boson mass of $M_H>114.4 \gev$~\cite{LEPHiggsSM}
from the LEP Higgs searches
does not apply generically. The same is true for the limits from
the searches for a SM-like Higgs at the \lhcee, which essentially rule out
the mass range $127 \gev \lsim \MH \lsim 600\gev$ for a SM-like Higgs.
As has been demonstrated in various benchmark scenarios 
in the MSSM~\cite{LEPHiggsMSSM} and the
NMSSM~\cite{Mahmoudi:2010xp}, allowed parameter regions for Higgs bosons
of those extended Higgs sectors exist both below the LEP limit and above 
$127 \gev$. In order to test whether a given point in the (N)MSSM
parameter space is allowed or ruled out by the Higgs searches at LEP,
the Tevatron, and the \lhcee\ one therefore needs to confront the
predictions of the model with the available cross section limits in the 
various search channels at each collider. For this purpose we make use
of the code {\tt HiggsBounds}~\cite{Bechtle:2011sb} (version
3.6.1-beta). In order to obtain the correct statistical interpretation
in terms of a 95\% CL exclusion limit, {\tt HiggsBounds} uses the input
provided for the model under consideration (in the case of our analysis 
effective couplings and partial widths of the Higgs bosons of the MSSM
and the NMSSM) to determine for each considered parameter point 
the channel that has the highest expected sensitivity for an exclusion. 
Only for this particular channel the
theory prediction is then compared to the observed experimental limit,
which determines in a statistically consistent manner whether the
parameter point is allowed or excluded at the 95\% CL.
The version of {\tt HiggsBounds} used for our analysis 
includes in particular limits
from CMS on an MSSM Higgs decay into a tau pair~\cite{CMStbmalimit},
which have a strong impact on the allowed region in the ($\MA,\,\tb$) plane.%
\footnote{Using the latest limits from the CMS 
update~\cite{CMStbmalimit_latest} in this channel would lead to only
minor changes in the results of our analysis.}%
~Results from $h\to \gamma\gamma$~\cite{ATLAS_gaga_Dec11,CMS_gaga_Dec11} and 
$h\to WW$ channels~\cite{ATLAS_WW_Dec11,CMS_WW_Dec11} (among others) are also
implemented.%
\footnote{Strictly speaking, these results can only be directly
interpreted in models with SM-like Higgs production. Even if this does
not hold exactly, we have verified for the scenarios considered below
that the ratio between the contributions from gluon fusion to vector
boson fusion do not deviate from the SM case
to the degree that vector boson fusion
becomes the dominant production mode. Since we are mainly interested
in investigating scenarios with an \emph{enhanced} rate, we take a
conservative approach and apply these bounds for maximal exclusion.}%
~Due to their importance to our analysis, we will also display these limits
separately below.

\subsubsection*{Bounds from direct searches for SUSY particles}

Constraints also exist from direct searches for supersymmetric particles
at LEP, the Tevatron, and the LHC. The least model-dependent limits
are the ones from LEP. In particular the limit on the lightest chargino
mass, $m_{\chi_1^\pm}>94 \gev$~\cite{pdg}, applies to both the MSSM and
the NMSSM and restricts the parameter $\mu$ ($\mueff$) of the (N)MSSM
to values above about $100 \gev$. For the squarks of the first two
generations the LHC mass limits are most stringent~\cite{susy11}, but
have a certain model dependence. The ATLAS and CMS collaborations set limits 
above $1 \tev$ in particular ``simplified models'' for the masses of first
and second generation squarks and for the gluino; slightly weaker bounds are
derived in the constrained MSSM. For the third generation squarks, which
are most important for the radiative 
corrections in the Higgs sector, the bounds are much weaker and more
model dependent. This means in particular that the presence of large
stop mixing (as favoured by a relatively high value of the Higgs mass of
about $125\gev$, see for instance \citere{Mh125our})
is not excluded at present.

\subsubsection*{The anomalous magnetic moment of the muon}

A significant deviation exists between the experimentally
measured value of the anomalous magnetic moment of the muon,
$a_{\mu}^{\mathrm{exp}}$ and the theoretical prediction in the
SM. Numerically this deviation amounts to $\De
a_\mu=a_\mu^{\mathrm{exp}}-a_\mu^{\mathrm{SM}}= (30.2 \pm 8.8)\times
10^{-10}$\cite{newDavier,newBNL}, which corresponds to more than a
$3\,\sigma$ effect. Employing new physics contributions to account
for this deviation leads to bounds on the model
parameters~\cite{Gunion:2005rw,Domingo:2008bb}. The
dominant contributions to $a_\mu$ in the NMSSM are known including
leading corrections up to the two-loop order~\cite{Domingo:2008bb}. 
For the numerical evaluation we use \NMSSMTools. 
As the $2\,\sigma$ allowed range for the 
NMSSM-specific contributions $\Delta
a_\mu^{\mathrm{NMSSM}}=a_\mu^{\mathrm{NMSSM}}-a_\mu^{\mathrm{SM}}$ we use 
$1.21\times 10^{-9}<\Delta a_\mu^{\mathrm{NMSSM}}<4.82\times 10^{-9}$,
which includes a theory uncertainty on the SUSY evaluation corresponding
to $2.0\times 10^{-10}$ added in quadrature to the uncertainty quoted
above. We note that, similarly to the $\mu$ parameter in
the MSSM, a positive value for $\mueff$ is strongly favoured when
$a_{\mu}$ is included as a constraint.

\subsubsection*{Flavour physics}
A recent analysis and summary of flavour physics constraints on the
NMSSM parameter space has been presented in~\cite{Mahmoudi:2010xp}. In
the present setup we use 
\NMSSMTools~(version 2.3.5) to evaluate the NMSSM theory
predictions. The corresponding experimental limits are
listed in Table~\ref{tab:flavour}. Parameter-dependent theory
uncertainties are added linearly to the intervals shown in the table
before evaluating exclusion. 

In theories with minimal flavour violation (MFV), which we are
investigating here, the strongest
constraints from flavour physics can usually be derived from
$B$-physics observables such as $\br(B\to X_s\ga)$,
$\br(B_s \to \mu^+\mu^-)$, 
$\br(B_u \to \tau^+\nu_\tau)$, or from the mass mixings $\De M_s$, 
$\De M_d$~\cite{Hiller,Domingo:2007dx}. In addition to the
MSSM-type contributions from sparticles and non SM-like Higgs
states (charged and neutral), 
the NMSSM may also be further constrained in the case
where a very light $\cp$-odd Higgs boson is 
present~\cite{Dobrescu:2000jt}.

It has been argued that NMSSM specific contributions, i.e.\
contributions that go beyond the MSSM, to
$\br(B\to X_s\ga)$ are negligible~\cite{Domingo:2007dx}. 
Besides the charged Higgs boson contribution, which typically
yields large effects for a relatively small $\MHp$,
we therefore include also the relevant
loops of charginos, gluinos, as well as (non-singlet) neutralinos.
Depending on the SUSY parameters, large cancellations between different 
contributions are possible, and any of the contributions mentioned above
can become dominant. If a specific benchmark scenario is chosen
for the SUSY parameters in the other sectors,
it cannot be expected that such a cancellation takes place for all values
of the Higgs sector parameters. However, a comparably small shift in the
underlying SUSY scenario would typically be sufficient to give
rise to large compensations between the different contributions, which
could bring the prediction into agreement with the experimental result.

Both the processes $\br(B_s \to \mu^+\mu^-)$ and
$\br(B_u\to \tau^+\nu_\tau)$ are also very sensitive to new
physics effects, and can mainly exclude regions of parameter space with
large $\tb$. But while unacceptably large contributions to
$\br(B_s \to \mu^+\mu^-)$ can be
avoided in a similar fashion as for $\br(B\to X_s\ga)$ by a slight change
of the other parameters of the SUSY scenario, such a compensation in
general does not work for $\br(B_u\to \tau^+\nu_\tau)$. This
is due to the fact that this process involves charged Higgs exchange 
at tree-level. As a consequence, this process provides a similar exclusion 
power in the ($\MHp$, $\tb$) plane for the MSSM and the NMSSM.

\begin{table}
\centering
\begin{tabular}{lcccc}
\hline
Observable & Exp.~lower limit & Exp.~upper limit \\
\hline
$\mathrm{BR}(B\to X_s\gamma)_{E_\gamma > 1.6\;\mathrm{GeV}}$ & $3.03\times 10^{-4}$ &$4.07\times 10^{-4}$ & \cite{Barberio:2006bi}\\
$\mathrm{BR}(B_s\to \mu^+\mu^-)$ & - & $1.1\times 10^{-8}$ & \cite{Bsmumu}\\
$\mathrm{BR}(B^\pm\to \tau\nu_\tau)$ & $0.79\times10^{-4}$  & $2.45\times10^{-4}$ & \cite{Asner:2010qj}\\
$\Delta M_{B_s}$ & $17.53$ ps$^{-1}$ & $18.01$ ps$^{-1}$ & \cite{DMBs}\\
$\Delta M_{B_d}$ & $0.499$ ps$^{-1}$ & $0.515$ ps$^{-1}$ & \cite{Barberio:2006bi}\\
\hline
\end{tabular}
\caption{Experimentally allowed ranges at the $2\,\sigma$ level used for
  the flavour physics observables.} 
\label{tab:flavour}
\end{table}


\section{Numerical analysis}
\label{sect:results}
In this section we analyse numerically the phenomenologically important 
loop-induced Higgs decays of the neutral $\cp$-even Higgs bosons
to two photons,
\begin{equation}
\Hi \to \ga\ga ~(i = 1,2,3)~. 
\end{equation}
For completeness, we also discuss the decays of the $\cp$-odd Higgs
bosons,
\begin{equation}
a_j \to \ga\ga ~(j = 1,2).
\end{equation}
We investigate in particular, taking into account the existing
constraints on the parameter space discussed above,
to what extent the phenomenology of Higgs decays into two photons 
can differ in the MSSM and the NMSSM from the SM case.
We choose a baseline MSSM benchmark scenario, fixing the soft SUSY-breaking parameters according to: 
\begin{align}
 \msusy = \MsqL = \MsqR  &= 1000 \gev, \non \\
  \MslL = \MslR &= 250\gev \mbox{ (to comply with $\De a_\mu$)}, \non \\
  A_t &= A_b = A_\tau, \non \\
\MOne &= 
\frac{5}{3} \tan^2 \thw \MTwo \approx \edz \MTwo, \non \\
 \MTwo &=400\gev \non \\
 \mgl &= 1200 \gev.
\label{eq:MSSMparams}
\end{align}
While in the MSSM the tree-level Higgs sector can be specified by the
two parameters $\MHp$ (or $\MA$) and $\tb$, 
the NMSSM Higgs sector has larger
freedom and requires additional input. We choose the following set of
parameters to describe a point in the NMSSM parameter space: 
\begin{align}
\MHp, \; \tb, \; \la, \; K \equiv\ka/\la, \; \Aka~.
\label{eq:higgsparam}
\end{align}
The parameter $\MHp$ here in principle plays the same role as in the
MSSM. However, since we employ \NMSSMTools\ to calculate the Higgs masses, the input $\MHp$ is not defined in the on-shell renormalisation scheme, 
and must be understood as a tree-level input mass which can
be directly translated  to a value for $\Ala$ using \refeqs{eq:MA2},
\eqref{eq:MHp2}. The calculated physical $\MHp$
(including the higher order corrections) will therefore in general not
be identical to the input value.%
\footnote{
This feature would be avoided with an on-shell renormalisation of $\MHp$,
see e.g.~\cite{Ender:2011qh,mhcMSSMlong}.}%
~The parameter  
$\mueff$, which we take as input, is in close
correspondence to the $\mu$ parameter in the MSSM. Together with
$\la$ it determines the value of the singlet vev
through~\refeq{def-mueff}.

Below we are going to perform scans over the NMSSM parameter space. For
the comparison between the NMSSM, the MSSM, and the SM we classify the resulting scenarios as follows: 
\begin{itemize}
\item[(i)] General NMSSM\\
This refers to any scenario in the NMSSM specified by the full set of
parameters given by \refeq{eq:higgsparam} and the MSSM parameters
in \refeq{eq:MSSMparams} (where we also defined our default
settings and values).
\item[(ii)] The MSSM limit\\
To recover the MSSM, we start from an arbitrary NMSSM scenario and take
\begin{align}
\la \to 0, \; \ka \to 0, \qquad K\equiv\ka/\la = \mathrm{constant}.
\label{eq:mssmlim}
\end{align}
All other parameters (including $\mueff$) are held fixed,
which corresponds to $v_s \to \infty$. In this limit the Higgs doublet
(and neutralino) couplings become MSSM-like, and the couplings of the
singlets vanish. It should be noted, however, that the decoupled
singlets do not necessarily 
correspond to the heaviest states $h_3$, $a_2$, and $\neu{5}$
even in this limit. In the MSSM limit the parameters $\MHp$ and $\tb$
correspond exactly to their MSSM counterparts. 

\item[(iii)] Decoupling (SM) limit\\
On top of the condition given by \refeq{eq:mssmlim} for the
MSSM limit, we define decoupling to the SM case
by taking the additional limit  
\begin{align}
\MHp\gg M_Z.
\end{align}
This leads to heavy, nearly mass-degenerate, doublet Higgs states and
leaves one light Higgs boson, $H$, with SM-like couplings to vector
bosons and fermions. The singlet states are not affected by this limit;
they remain decoupled with finite masses (which may or may not be lower than
$M_H$). 

\item[(iv)] SM+singlet limit\\ 
When the doublet decoupling condition $\MHp\gg M_Z$ is fulfilled for
points with finite non-zero $\la$, $\ka$ (i.e., values that differ
from the MSSM
limit) we speak of the SM+singlet limit. This name is appropriate, since
the low mass Higgs spectrum in this scenario consists of two $\cp$-even and
one $\cp$-odd degree of freedom. 
\end{itemize}


\subsection{Decays of $\cp$-even Higgs bosons in the MSSM}
\label{sect:MSSM}
Before we proceed to the NMSSM case, we study the two photon decays
of the two $\cp$-even Higgs bosons, $h$ and $H$, in the MSSM and compare
to the SM.   
For the numerical evaluation in the MSSM we use the code 
{\tt FeynHiggs} (version
2.8.6)~\cite{feynhiggs,mhiggslong,mhiggsAEC,mhcMSSMlong}, which is also
used to evaluate SM quantities given in this section.  
To study interesting regions of the MSSM parameter space, where
differences in the di-photon channel between the MSSM and the SM can
occur, we perform 
a random scan over the parameter ranges given in \refta{tab:hrangesMSSM}.
The remaining MSSM parameters are kept at their `benchmark' values
specified in \refeq{eq:MSSMparams}.

It should be noted that we allow for comparably high values for
$\mu$; this is relevant for the possible size of some of the effects 
that we will discuss in detail below.
However, such large values of $\mu$, together with large values of $\tb$,
can lead to parameter combinations that show a non-stable behaviour in
perturbation theory. In order to avoid parameter combinations that
result in unacceptably
large two-loop corrections in the evaluation of the Higgs
boson self-energies and related quantities, we implement 
an upper limit on the corrections to the elements of the
$\matr{Z}$~matrix (see \citere{mhcMSSMlong}). Comparing
the one- and two-loop values of the respective diagonal elements,
we require the following condition for the light $\cp$-even Higgs,
$||Z_{11}^{\rm 2-loop}| - |Z_{11}^{\rm 1-loop}||/|Z_{11}^{\rm 1-loop}| < 0.25$, 
and analogously for the heavy $\cp$-even Higgs with the replacement 
$Z_{11} \to Z_{22}$.
We found that this upper bound is effective for avoiding parameter regions
that are deemed unstable under higher-order corrections.

\begin{table}
\centering
\begin{tabular}{ccc}
\hline
Parameter & Minimum & Maximum \\
\hline
$\msusy$ & $750$ & $1500$ \\
$M_2\simeq 2M_1$ & $200$ & $500$ \\
$A_t=A_b=A_\tau$ & $-2400$ & $2400$ \\
$\mu$ & $200$ & $3000$ \\
\hline
$\MA$ & $100$ & $600$ \\
$\tb$ & $1$ & $60$ \\
\hline
\end{tabular}
\caption{Parameter ranges for the MSSM scan. All parameters with mass
  dimension are given in GeV.} 
\label{tab:hrangesMSSM}
\end{table}

\begin{figure}[htb]
\centering
\includegraphics[width=0.48\columnwidth]{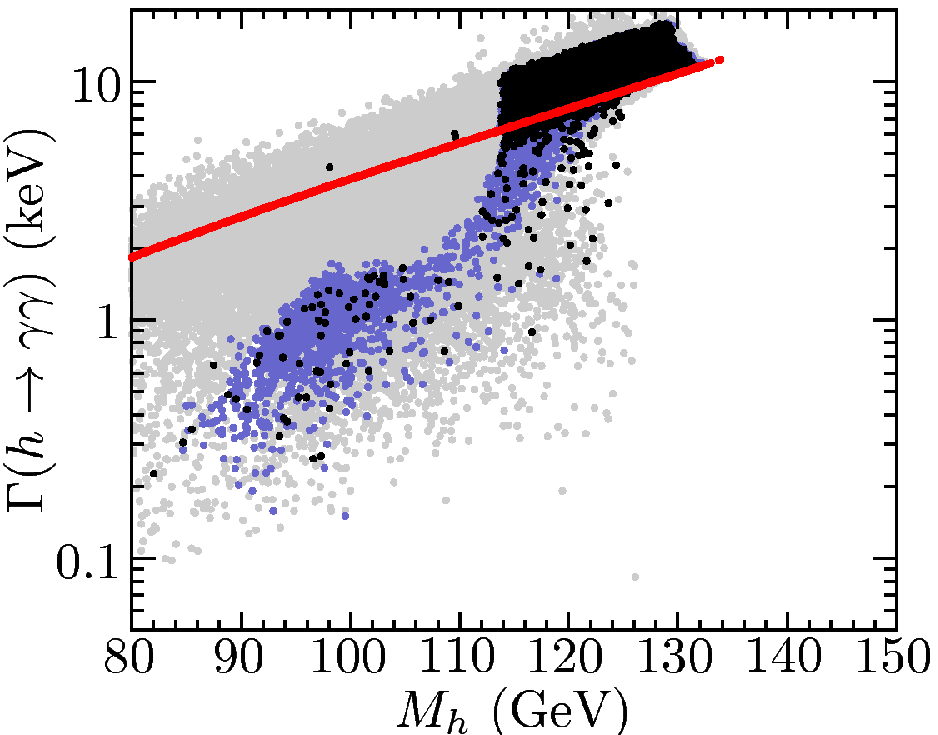}
\includegraphics[width=0.48\columnwidth]{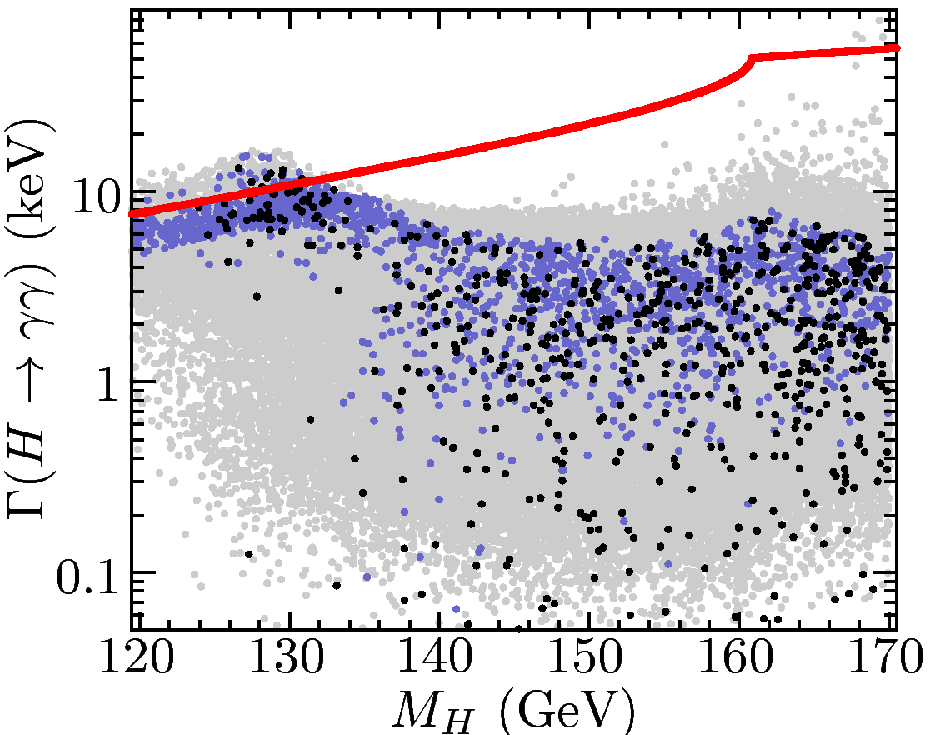}\\
\includegraphics[width=0.48\columnwidth]{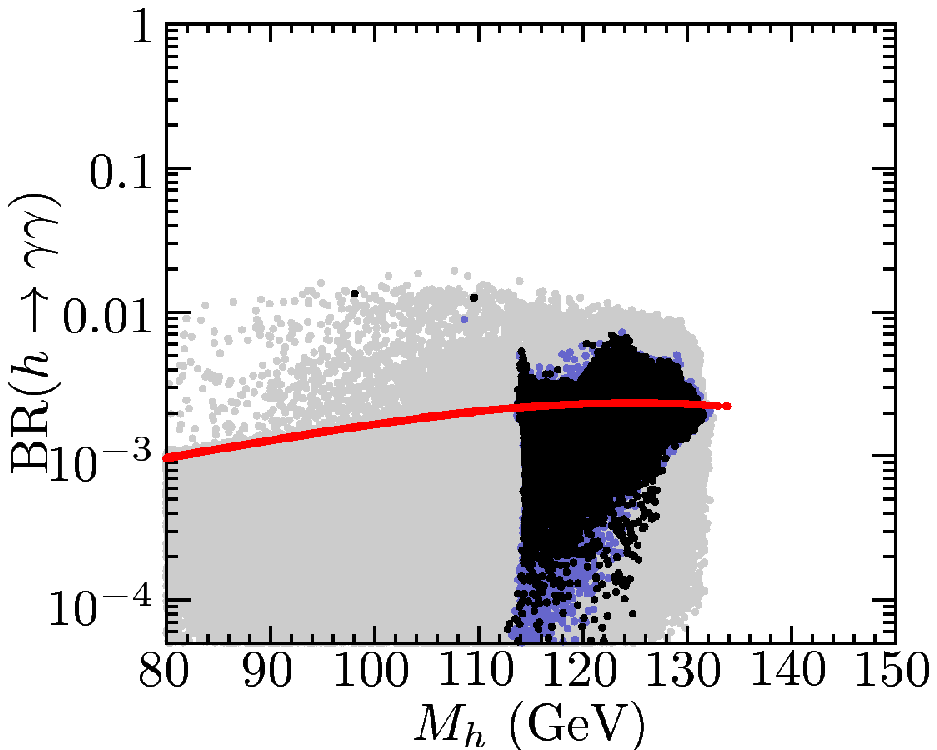}
\includegraphics[width=0.48\columnwidth]{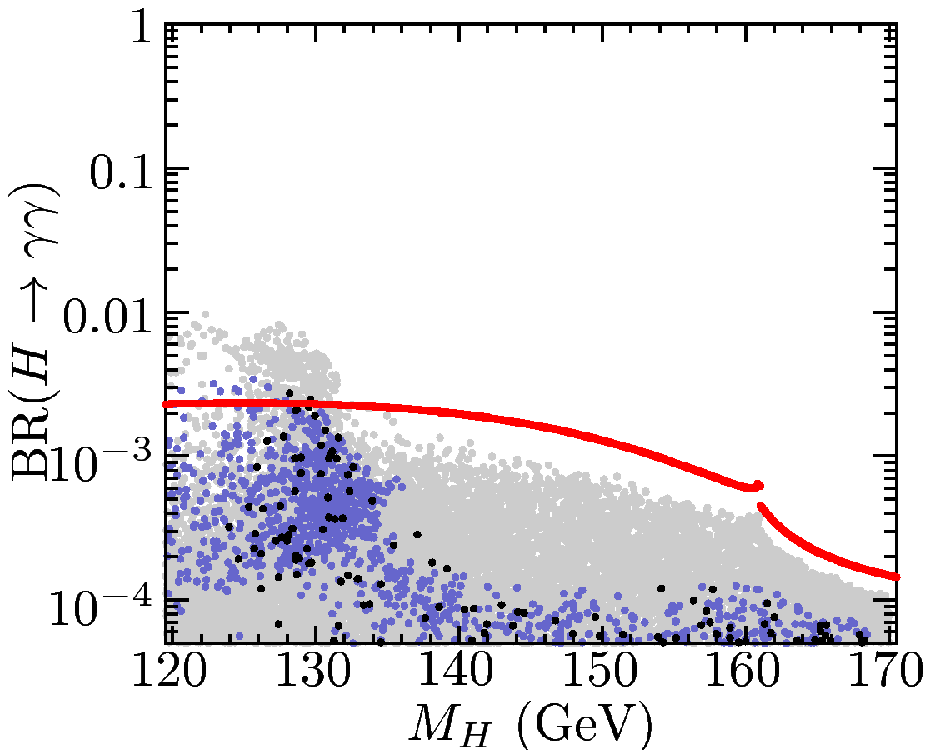}
\caption{Results from the MSSM parameter scan for the partial widths
  $\Ga(h,H\to\ga\ga)$ of $h$ (left) and $H$ (right), and the
  corresponding branching ratios. The full result of the scan 
  (all points allowed by the theoretical constraints
and the direct search limits for sparticles) is shown
  in grey. The blue points are compatible with the direct Higgs search
  limits (from {\tt HiggsBounds}~3.6.1, i.e.\ including \lhcee),
  while the black points in addition give a result in agreement 
  with $(g-2)_\mu$ and $\mathrm{BR}(b\to s\gamma)$. The solid (red)
  curve shows the respective quantities  
 evaluated in the SM. 
}
\label{fig:rgaga_mssm_br}
\end{figure}

In \reffi{fig:rgaga_mssm_br} we show $\Ga(h \to \ga\ga)$ in the top left
and $\br(h \to \ga\ga)$ in the bottom left plot as a function of $\Mh$. 
The corresponding plots
for $H \to \ga\ga$ are given in the right column.
The colour coding is as follows:  
all points in the scan which are allowed by the theoretical constraints
and the direct search limits for sparticles~\cite{pdg}, as discussed
above, are plotted in grey. 
Points which are also allowed by direct Higgs search limits (from
{\tt HiggsBounds}~3.6.1, \htg{i.e.\ including \lhcee}) are shown in 
blue (on top of the grey points). 
Finally, points which fulfil additionally 
the constraint from $(g-2)_\mu$ and $\br(b\to s\gamma)$ (both
are here calculated with \FeynHiggs) are plotted in black. 
The red (solid) curve 
in \reffi{fig:rgaga_mssm_br} shows the corresponding SM
result with $\MHSM$ set equal to the corresponding MSSM Higgs mass. 
It should be noted that here (and in all
the following plots) different densities
of points appearing in different regions have no physical meaning, as 
the point density is related to the specific procedure chosen for
the sampling of the SUSY parameter space.

We first focus on the light $\cp$-even Higgs boson, $h$, decaying into
two photons.  
The extra particles in the MSSM yield additional loop
contributions, which can both lower and raise $\Gamma (h \to \ga\ga)$ 
compared to the SM case. 
For $\Mh < 114.4 \gev$%
\footnote{\htg{
We neglect here, and in the following,
the theory uncertainty of the Higgs 
boson mass evaluation, which for the light Higgs boson should be 
roughly at the level of $2-3 \gev$~\cite{mhiggsAEC}.}}%
~most of the scenarios where $\Ga(h \to \ga\ga)>\Ga(\HSM \to \ga\ga)$
are ruled out by the direct Higgs search limits, but we 
also find allowed points in this region. For those $h$ 
couples with about SM strength to gauge bosons, but is
nevertheless not excluded due to a (much) suppressed coupling to $b$
quarks, which weakens the corresponding LEP limit. The fact that an
enhanced rate for the decay of a Higgs boson to two photons is possible
even below the LEP limit provides
a motivation to extend the LHC
Higgs searches to this region.
In the following we focus on the mass region above the
LEP limit. There we find scenarios in which $\Ga(h \to \ga\ga)$ 
is enhanced by up to $\sim 70 \%$ with respect to the SM. 
On the other hand, as can be seen from the lower left plot in
\reffi{fig:rgaga_mssm_br}, the $\br(h\to \ga\ga)$ can be
enhanced by a factor $\sim 3$ over the SM in the same mass range
(due to a suppression of the $b \bar b$ decay mode as discussed
in more detail
below). It is interesting to note that for the points that are allowed 
by all constraints 
the maximum enhancement of the branching ratio occurs around $\Mh \sim
125\gev$.
\htg{One can finally observe that no SM values are reached for 
$\Mh \lsim 114.4 \gev$, reflecting the fact that a SM-like Higgs boson
is ruled out by the LEP Higgs searches.}

The corresponding results for the heavy $\cp$-even MSSM Higgs
boson are shown
in the right column of
\reffi{fig:rgaga_mssm_br}. For $\MH \lsim 130 \gev$ we find viable
points with a BR slightly larger than for a SM Higgs boson.
For larger values of
$\MH$ one can see the behaviour expected from the decoupling
properties of the MSSM, i.e.\
$\Ga(H \to \ga\ga)$ and $\br(H \to \ga\ga)$ are
both suppressed with respect to the SM, with the level of suppression
increasing with $\MH$.

In order to investigate the phenomenology at the LHC, besides the 
branching ratio of course also the Higgs production cross section ---
which in general will also be modified in a model of physics beyond the
SM --- 
has to be taken into account. The combined enhancement or
suppression over the SM for a process $pp\to \Hi \to X$ can therefore 
be summarised in the ratio 
\begin{align}
R_{X}^{\Hi}=\frac{\sigma(pp\to \Hi)\times \br(\Hi\to X)}
                {\sigma(pp\to \HSM)\times \br(\HSM\to X)}~. 
\label{eq:Rgaga_full}
\end{align}
In the MSSM $\Hi$ denotes either $h$ or $H$, while in the NMSSM (discussed 
in the next subsection) $\Hi$ ($i=1\ldots 3$) can be any of
the three $\cp$-even Higgs states. 
If the Higgs production cross section is dominated by a single
mechanism, such as gluon 
fusion which is often the case at the LHC%
\footnote{
We have checked that for the relevant regions of parameter space
discussed below the gluon
fusion production cross section always strongly dominates over the associated
Higgs boson production from bottom quarks.}%
, a common approximation is to
use instead of $\si(pp\to \Hi)$ the parton-level cross section
$\hat{\si}(gg\to \Hi)$. Neglecting the differences in kinematics,
the decay width $\Ga(\Hi\to gg)$ has the same dependence as
$\hat{\si}(gg\to \Hi)$ on the couplings of the involved particles,
and the dominant higher-order QCD corrections are expected
to cancel out in the ratio.%
\footnote{
Non-negligible differences are mainly expected if the bottom loop
contribution to $\Hi\to gg$ dominates over the top loop
contribution. In the case of 
the light $\cp$-even Higgs boson can happen for very low $\MA$
and moderate to large $\tb$ values, whereas in the case of the
heavy $\cp$-even Higgs boson this can happen for larger $\MA$ and 
$\tb \gsim 5$.
Our results therefore exhibit an
additional uncertainly in this part of the parameter space.
Additional loop contributions from SUSY particles, while taken into
account in our calculation, are usually subdominant and of lesser
importance in this context.}%
~Making use of this approximation, \refeq{eq:Rgaga_full}
can be expressed as 
\begin{align}
R_{X}^{\Hi} \simeq 
  \frac{\Ga(\Hi\to gg)\times \br(\Hi\to X)}
       {\Ga(\HSM\to gg)\times \br(\HSM\to X)} 
= \frac{\Ga(\Hi\to gg)\times \Ga(\Hi\to X)
                      \times \Ga_{\mathrm{tot}}(\HSM)}
       {\Ga(\HSM\to gg)\times \Ga(\HSM\to X)
                      \times \Ga_{\mathrm{tot}}(\Hi)}. 
\label{eq:Rgaga}
\end{align}
This definition will be used to calculate $R_{\ga\ga}^{\Hi}$ and
$R_{WW}^{\Hi}$ in the MSSM (and also in the NMSSM below). 

\begin{figure}
\centering
\includegraphics[width=0.48\columnwidth]{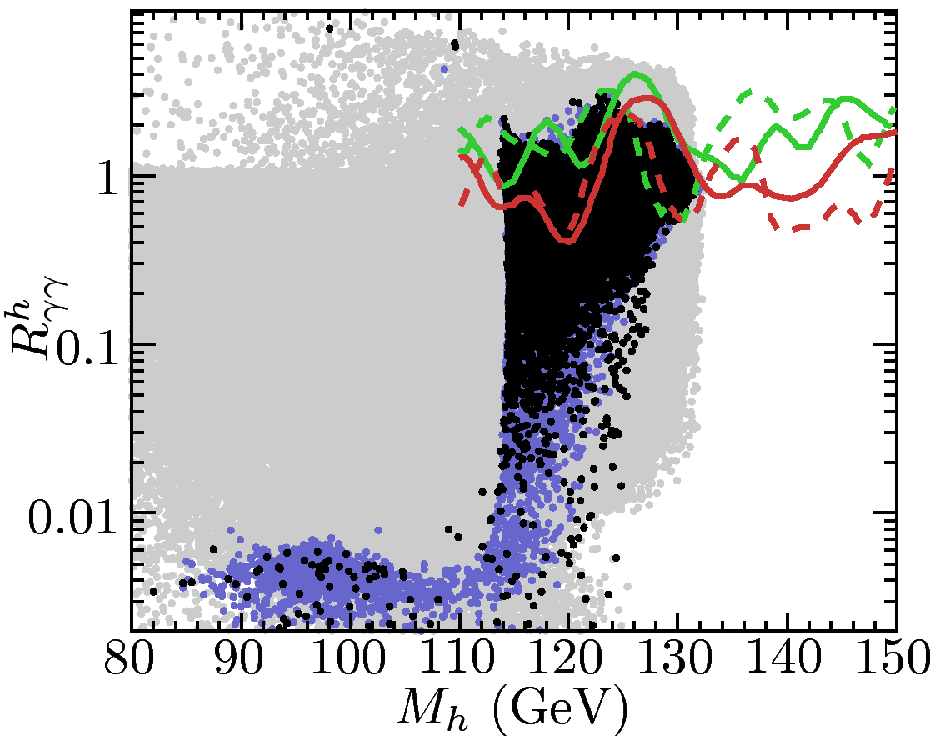}
\includegraphics[width=0.48\columnwidth]{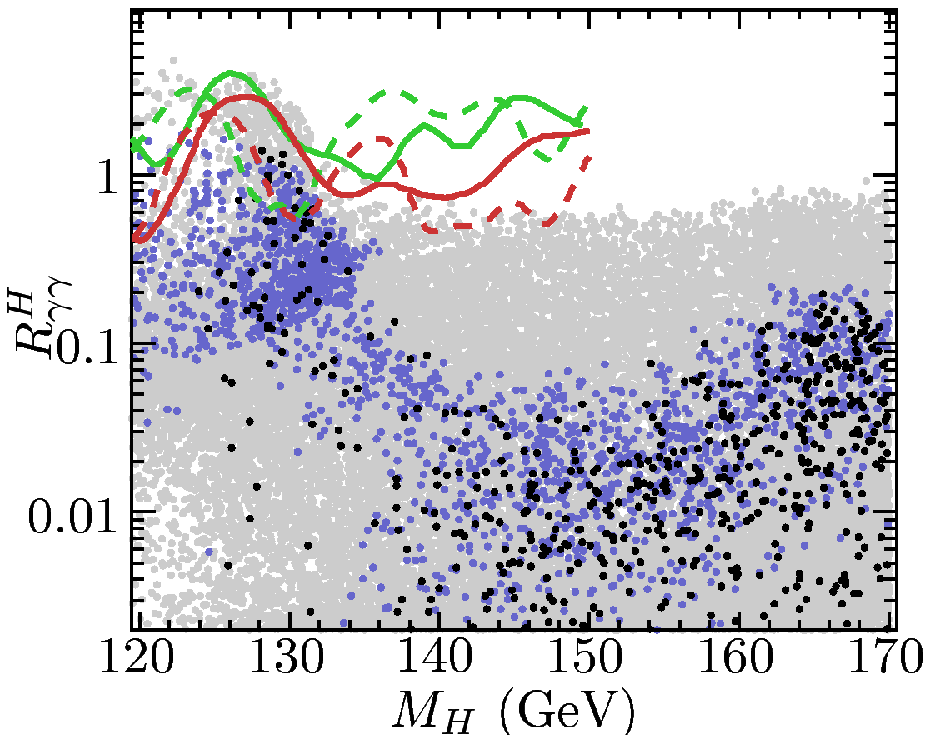}
\includegraphics[width=0.48\columnwidth]{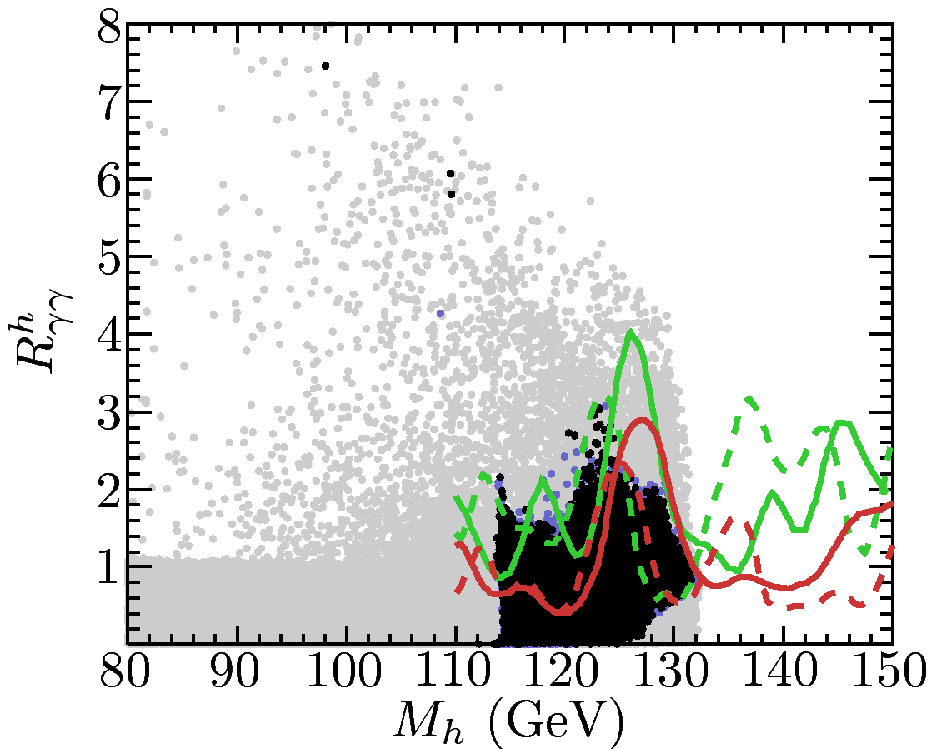}
\includegraphics[width=0.48\columnwidth]{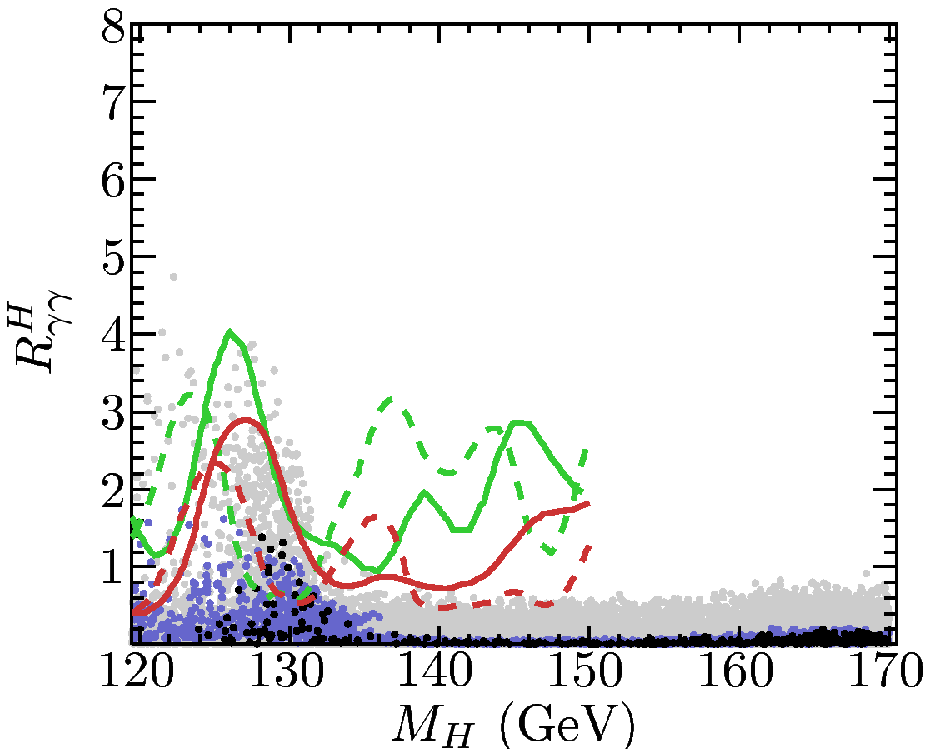}
\caption{Results from the MSSM parameter scan on the ratios
  $R_{\ga\ga}^{h}$ for the light $\cp$-even Higgs boson $h$ (left
column) and $R_{\ga\ga}^{H}$ for the heavy $\cp$-even Higgs boson $H$
(right column). The plots are displayed both on a logarithmic scale
(upper row) and on a linear scale (lower row).
  The colour coding for the scan points is the same as in
 \reffi{fig:rgaga_mssm_br}.
  The green lines are the corresponding limits from \htr{data presented
   in 2011 data from}
  ATLAS~\cite{ATLAS_gaga_Dec11} (solid) and from
  CMS~\cite{CMS_gaga_Dec11} (dashed). 
  \htg{The red lines are the new limits from ATLAS (solid) and CMS (dashed)
    taken from~\cite{discovery}.}
}
\label{fig:rgaga_mssm}
\end{figure}

The results for $R_{\ga \ga}^{h}$ and $R_{\ga\ga}^{H}$ are shown in
\reffi{fig:rgaga_mssm}, with the same 
colour coding as in \reffi{fig:rgaga_mssm_br}.
In order to make the results better
visible we display them twice on a logarithmic scale
(upper row) and on a linear scale (lower row). 
The green curves in \reffi{fig:rgaga_mssm} 
show exclusion
limits in the di-photon channel at $95\%$ CL from 
\htr{data presented in 2011 data from} 
ATLAS~\cite{ATLAS_gaga_Dec11} (solid) and CMS~\cite{CMS_gaga_Dec11}
(dashed).\footnote{\htg{The 2011 exclusion limit from CMS in 
this channel was updated in \cite{gaga_latest}. Including} 
this limit in our analysis 
would not qualitatively change our results.}
\htg{The red lines are the new limits from ATLAS (solid) and CMS (dashed)
taken from~\cite{discovery}.}
The exclusion limits from ATLAS and CMS are displayed here explicitly 
for comparison, but \htg{only the \lhcee\ data enters} 
our analysis also as part of the 
constraints implemented in {\tt HiggsBounds}. As explained above, 
{\tt HiggsBounds} considers only the single channel with the 
highest \emph{expected} sensitivity for determining 95\% CL (combined)
exclusion. In the considered region the expected sensitivity of the CMS
search~\cite{CMS_gaga_Dec11} happens to be slightly higher than the one
from ATLAS~\cite{ATLAS_gaga_Dec11}, so that 
only the CMS limit actually has an effect in our analysis. The plot
shows also some allowed points with $R_{\ga\ga}^{h}$ above the CMS \htg{2011}
exclusion curve. For these points another channel has a higher expected
sensitivity, so that the $\ga\ga$ channel has not been selected by 
{\tt HiggsBounds} for determining the 95\% CL limit. 

As one can see in the left column of \reffi{fig:rgaga_mssm}, 
for $R_{\ga \ga}^{h}$ in principle a large enhancement, roughly up to a
factor six, would be possible in the mass range $\Mh = 114 \ldots 130 \gev$ 
(and an even stronger enhancement for lighter masses). Such large
enhancements are now ruled out by the LHC searches in the $\ga\ga$
channel. For the points that are allowed by all the considered
constraints we find that in the region above the LEP limit 
a suppression of $R_{\ga \ga}^{h}$ 
by more than an order of magnitude is possible. 
A maximal enhancement of about $\sim$ 3 times
the SM value, on the other hand,  occurs for $\Mh\approx124 \gev$. 
\htg{This is in interesting agreement with the recent data announcing a
  discovery of a new state compatible with a Higgs boson
close to $\sim 125\gev$,} that has been \htg{reported} both by 
ATLAS and CMS. This \htg{observed} excess is
compatible with a SM Higgs signal, or \htg{even better with} a signal of
\htg{a somewhat} enhanced strength in the two photon channel. 
Our results show that the MSSM could account for an enhanced Higgs
signal as compared to the SM case around $\Mh=125\gev$ with the maximal
strength that is allowed by the present limits from ATLAS and CMS. 
The detailed origin of this enhancement will be discussed below. 
\reffi{fig:rgaga_mssm} also shows that 
the possible size of the enhancement decreases for
larger $\Mh$; 
for $\Mh = 130 \gev$ $R_{\ga \ga}^{h}$ is confined to values 
close to unity for the allowed points in the parameter space.

The right column of \reffi{fig:rgaga_mssm} shows the 
corresponding results for the heavy $\cp$-even Higgs.
For $\MH \lsim 130 \gev$ the results for the heavy MSSM Higgs
are qualitatively similar to the ones for the light $\cp$-even
Higgs. In particular, also in this case a slight enhancement over the SM
rate is possible for $\MH\approx 125\gev$ for the scan points that are
in agreement with the collider constraints (as discussed above, the
agreement with $(g-2)_\mu$ and the observables in the flavour sector
could be improved by modifying some of the SUSY parameters that do not
directly influence Higgs phenomenology). The size of the possible
enhancement turns out to have some sensitivity to the condition that we
have imposed on the elements of the $\matr{Z}$~matrix to ensure
perturbatively reliable results, see above. Somewhat larger enhancements
would be possible if the upper limit that we have imposed on the
  relative size of the two-loop corrections were relaxed. 
Our results for $R_{\ga\ga}^{H}$ demonstrate that the \htg{discovery of a new
boson in the}
$\ga\ga$ channel at a mass of about $125\gev$ that was observed by ATLAS
and CMS could also be interpreted within the MSSM as arising from the
\emph{heavier} $\cp$-even Higgs boson, as discussed in~\cite{Mh125our}.
Such a scenario would imply that besides a possible signal at about 
$125\gev$ there would be a lighter Higgs in the spectrum, having
significantly suppressed couplings to gauge bosons.
For $\MH \gsim 135 \gev$ we always find $R_{\ga\ga}^H < 1$, in
accordance with the decoupling properties of the MSSM.

Before turning to a discussion of the specific mechanisms responsible for 
possible enhancements of the di-photon channel in the MSSM as
compared to the SM case, we would like to make a brief comparison of our 
MSSM results to those existing in the literature. 
A previous study of the di-photon channel in the
MSSM~\cite{Cao:2011pg} found a smaller maximum enhancement,
$R_{\ga\ga}^{h}\sim 1.5$, and only for a very specific Higgs mass
region. We find a larger enhancement, primarily due to a larger scan range in
$\mu$. Restricting our scan ranges accordingly, our results agree with
those of \cite{Cao:2011pg}. Similarly, the analysis of \cite{1203.3207} uses 
a more restricted range of $\mu<1\tev$, and we also find agreement 
with our results for $R_{\ga\ga}^h$ in this parameter region.
In \cite{slac125}, it was claimed that a light Higgs boson with
$\Mh \approx 125 \gev$ and $R_{\ga\ga}^h\gtrsim 1$ cannot be realised in the
MSSM. Clearly, the results from our scan do not corroborate these
conclusions since we find that such a Higgs boson can indeed be realised
in the MSSM, possibly even with a rate 
\emph{enhanced} compared to the SM.

The issue of a possible enhancement of $R_{\ga\ga}^{h}$ for a Higgs
mass around $125\gev$ has also been discussed in
\cite{Mh125gaga}, where in particular the contributions from light staus
to $\br(h\to \ga\ga)$ and the suppression of $h\to b\bar{b}$ due
to Higgs mixing effects have been emphasised.
As seen above, we find that $\Ga(h \to \ga\ga)$ can exceed its SM value, 
which is found to be an effect of the stau loop contributions. 
The most sizeable enhancements observed in $R_{\ga\ga}^{h}$,
however, mainly arise from a suppression of the total width,
which in the SM is dominated by the partial decay width into 
$b \bar b$.
Suppressing the $b \bar b$
channel can therefore yield a significant reduction of
the total MSSM width. 
Such a suppression can happen in two different ways. The
reduced $hb\bar{b}$ coupling in the MSSM is given at tree-level by 
\begin{equation}
\frac{g_{hb\bar{b}}}{g_{\HSM b\bar{b}}} = -\frac{\sin \alpha}{\cos \beta},
\label{eq:hbb_tree}
\end{equation}
where $\alpha$ is the mixing angle in the $\cp$-even Higgs sector. 
In the decoupling limit $(\MA \gg \MZ)$ the SM is recovered,
i.e.~$(-\Sa/\Cb) \to 1$.   
Higher-order contributions from Higgs propagator corrections can
approximately be included via the introduction of an effective 
mixing angle, corresponding to the replacement  
$\al \to \aeff$~\cite{hff} (in our numerical analysis we treat
propagator-type corrections of the external Higgs bosons in a more
complete way, which is based on wave function normalisation factors that
form the $\matr{Z}$~matrix~\cite{mhcMSSMlong}). 
A suppression of the $h\to b \bar b$ channel thus occurs for small
$\aeff$. This feature motivated the definition of the 
``small $\aeff$'' benchmark scenario, see~\cite{smallalphaeff}.
In this scenario, the suppression of $\Gamma(h\to b\bar{b})$ occurs
for large $\tan \beta$ 
and very small values of $\MA$, below $200\gev$.

\begin{figure}[tb]
\centering
\includegraphics[width=0.48\columnwidth]{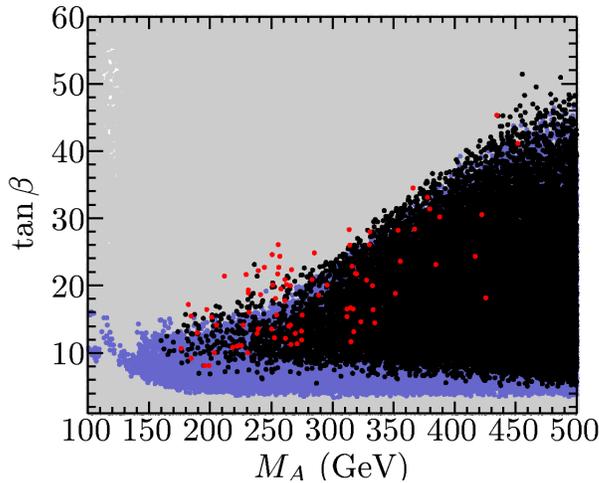}
\caption{Results from the MSSM parameter scan on the ratio
  $R_{\ga\ga}^{h}$ in the $(\MA,\tb)$ plane. The grey points are 
  excluded by the limits from direct Higgs searches
  \htg{(from {\tt HiggsBounds}~3.6.1, i.e.\ including \lhcee)}.
  MSSM Points with an enhancement of $R^h_{\ga\ga}$, corresponding to 
$R^{h}_{\ga\ga}>1$, are indicated in black, points with 
$R^{h}_{\ga\ga}>2$ are shown in red.}
\label{fig:tbma}
\end{figure}

Genuine corrections to the $hb\bar b$ vertex can lead to another type
of suppression. Beyond leading order, loop-induced Yukawa couplings of
$b$ quarks to the ``wrong'' Higgs doublet $H_{u}$ are induced. 
The modified $hb\bar{b}$ coupling can then be expressed as
\begin{equation}
\begin{split}
\frac{g_{hb\bar{b}}}{g_{\HSM b\bar{b}}} = \frac{1}{1+\De_b} \left( -\frac{\sin \aeff}{\cos \beta}+\De_b \frac{\cos \aeff}{\sin \beta}\right).
\label{eq:hbbdeltamb}
\end{split}
\end{equation}
Via the quantity $\db$~\cite{deltab1,deltab2} terms of
\order{(\als\tb)^n} 
and \order{(\alt\tb)^n} can be resummed.
The most relevant contributions are given by 
\begin{align}
\db &= \frac{2\als(\mt)}{3\pi} \, \tb \, \mgl \, \mu \,
                          I(\msbe^2, \msbz^2, \mgl^2) \;
     + \frac{\alt(\mt)}{4\pi} \, \tb \, \At \, \mu \, 
                          I(\mste^2, \mstz^2, |\mu|^2),
\label{def-db}
\end{align}
with
\begin{align}
I(a, b, c) &= - \frac{a b \ln(b/a) + a c \ln(a/c) + b c \ln(c/b)}
                 {(a - c) (c - b) (b - a)}~.
\end{align}
The dominant higher-order contribution to $\De_b$ are the QCD
corrections, given 
in~\cite{db2l}. Those contributions are not included in our analysis, 
but their effect can be approximated by using a scale of 
$\mt$ for the evaluation of the
one-loop expression, \refeq{def-db}.
While the loop-corrected $hb\bar{b}$ coupling, \refeq{eq:hbbdeltamb},
approaches the tree-level coupling, \refeq{eq:hbb_tree}, in the
decoupling limit ($\MA \gg \MZ$), a suppression of $g_{hb\bar{b}}$ is
possible for not too large $\MA$ if $\Delta_b$ is numerically sizable
and positive. For $\mu>1 \tev$ we find enhancements of $R_{\ga\ga}^{h}$
of more than 1.5 for values of $\MA$ up to roughly $450\gev$ and with
moderate to large values of $\tb$. 
Points with $R_{\ga \ga}^{h}>2$ are possible if $\De_b$ is relatively
large, $\Delta_b\simeq 0.5$.
The corresponding effect on $R_{\ga\ga}^h$ can be seen in
\reffi{fig:tbma} for the $(\MA,\tb)$ plane. 
The points with $R^h_{\ga\ga}>1$ are
indicated in black, and the ones with $R^h_{\ga\ga}>2$ are shown in red. 
The regions with only grey points are excluded by the limits from the
Higgs searches at LEP and the \htg{\lhcee}, where the latter ones arise from the
searches for MSSM Higgs bosons in the $\tau^+\tau^-$
channel~\cite{CMStbmalimit}. As can be seen from the figure, a slight
enhancement of $R^h_{\ga\ga}>1$ can be accommodated over the whole
allowed region displayed in \reffi{fig:tbma}, while scenarios with
$R^h_{\ga\ga}>2$ tend to be closer to the boundary of the region allowed
by the LHC Higgs limits.

\medskip
We now turn to the alternative case where $h$ is light and has
suppressed couplings to gauge bosons, 
whereas the heavier $\cp$-even Higgs $H$ is a SM-like Higgs boson. 
One finds a similar enhancement  
for $R_{\ga \ga}^{H}$, which is due to the suppression of
$g_{Hb\bar{b}}$, if 
\begin{align}
\frac{g_{Hb\bar b}}{g_{\HSM b\bar b}} = 
\frac{\cos \alpha}{\cos \beta}
\end{align}
is small. Such an enhancement is restricted to the mass region 
$M_H \lsim 130\gev$, since for higher mass values the coupling of the
heavy $\cp$-even Higgs to gauge bosons is suppressed, so that the
partial width $\Gamma(H \to \ga\ga)$ is smaller than for the SM case,
see \reffi{fig:rgaga_mssm_br}.
Accordingly, the scenarios with $R^H_{\ga\ga}>1$ are only realised in a 
relatively small parameter region close to the exclusion bounds from
the Higgs searches, for $\MA \lsim 150 \GeV$ and intermediate $\tb$.
The scenario in which $R_{\ga \ga}^{H}$ is enhanced is 
complementary to the one giving an enhancement in  $R_{\ga \ga}^{h}$
(as we checked explicitly). Consequently, 
a simultaneous enhancement in the di-photon channel for both $\cp$-even
Higgs bosons is not possible.  

\medskip
\begin{figure}[tb]
\centering
\includegraphics[width=0.48\columnwidth]{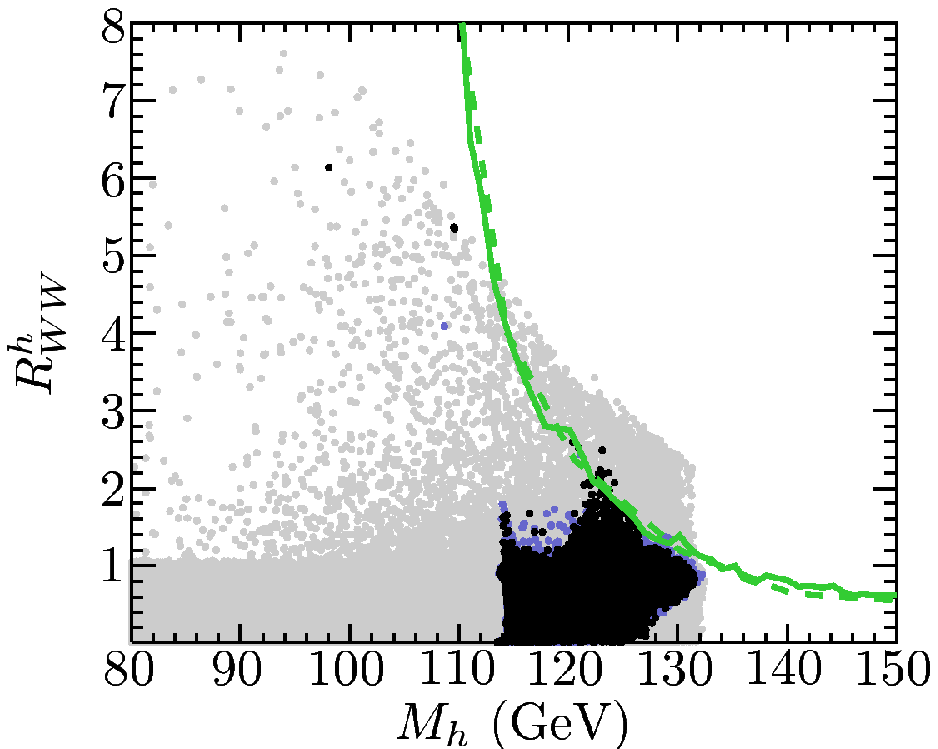}
\includegraphics[width=0.48\columnwidth]{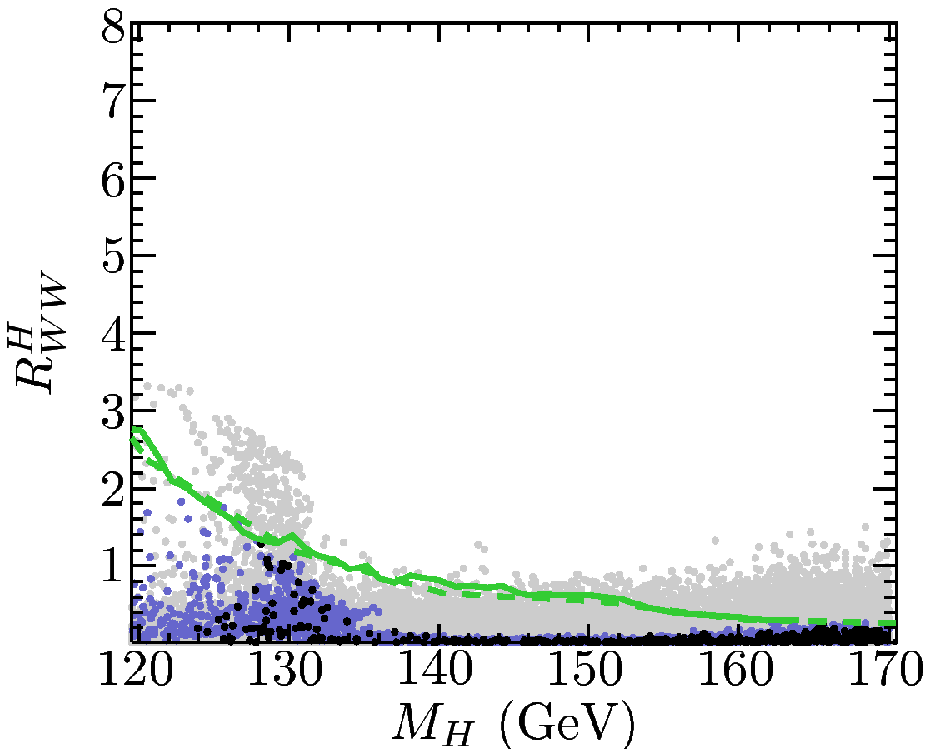}\\[.5em]
\includegraphics[width=0.48\columnwidth]{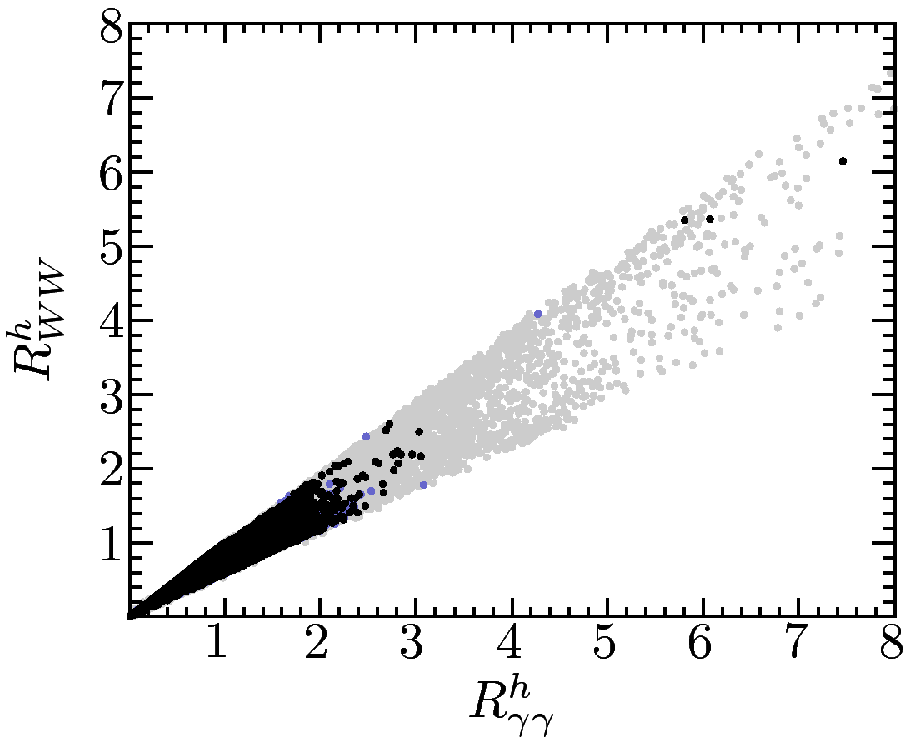}
\includegraphics[width=0.48\columnwidth]{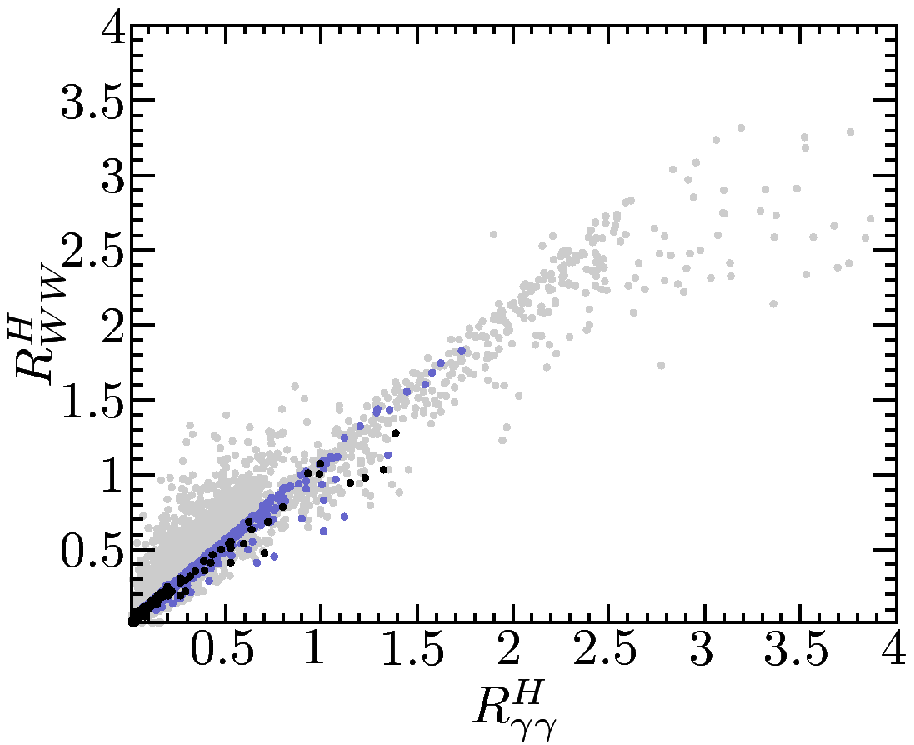} 
\caption{Results from the MSSM parameter scan on the ratios
$R_{WW}^{h}$ for the light $\cp$-even Higgs bosons $h$ (left column) and
$R_{WW}^{H}$ for the heavy $\cp$-even Higgs boson $H$ (right column).
The plots in the upper row show $R_{WW}^{h}$ and $R_{WW}^{H}$ as a
function of the respective Higgs mass. The $95\%$ CL 
exclusion limits for the $WW^{(*)}$ channel \htg{using \lhcee\ data}
from ATLAS \cite{ATLAS_WW_Dec11} 
(solid line) and CMS \cite{CMS_WW_Dec11} (dashed line) are also shown. 
The plots in the lower row show the correlation to $R_{\ga\ga}^{h}$ and 
$R_{\ga\ga}^{H}$.
The colour coding is the same as in \reffi{fig:rgaga_mssm_br}.}
\label{fig:RWW_mssm}
\end{figure}

A reduction of the total width, by the suppression of the 
$h,H \to b \bar{b}$
channel, can also affect the search for the Higgs boson in
other channels.  
In \reffi{fig:RWW_mssm} we investigate
the decays $h, H \to WW^{(*)}$. The colour
coding is as in \reffi{fig:rgaga_mssm_br}, but
here the green lines show the $95\%$ CL exclusion limits for 
the \htr{data presented in 2011 for the}
$WW^{(*)}$ channel from ATLAS \cite{ATLAS_WW_Dec11} (solid)
and CMS \cite{CMS_WW_Dec11} (dashed).
\htg{Here we do not show new results from~\cite{discovery}, since this
channel was updated only by CMS, and no combination of 2011 and 2012
data is available. The 2012 exclusion curve of CMS is very similar to
the one based on 2011 data.}
Similarly to $R_{\ga \ga}^{h}$, also $R_{WW}^{h}$ can be enhanced in the
MSSM. The maximal enhancement of $R_{WW}^{h}\approx 2.5$ that is
compatible with the \htg{existing} bounds is found
for $\Mh\approx125\gev$. These points, above the green lines, are
not excluded by the $h \to WW^{(*)}$ limits, because another channel
has a higher expected sensitivity for those points
(and consequently the $WW^{(*)}$
channel has not been selected by {\tt HiggsBounds} for determining the
95\% CL limit \htg{-- which would not change by the inclusion of the
  2012 data shown in~\cite{discovery}}).
Since the main contribution to $h \to \ga\ga$ comes from diagrams
containing $W$-boson loops, a strong correlation between these two
channels is expected. This correlation is confirmed by plots in
the lower row of \reffi{fig:RWW_mssm}. 
Contrary to $\Ga(h\to \ga \ga)$, $\Ga(h \to WW^{(*)})$ 
does not exceed its SM value, and therefore the maximal 
enhancement of $R_{WW}^{h}$ is always smaller than the enhancement of
$R_{\ga \ga}^{h}$. This leaves some room to have e.g.~a SM-like rate,
or even a rate that is somewhat suppressed w.r.t.\ the SM,
for $h\to WW^{(*)}$ (and $h\to ZZ^{(*)}$) with a simultaneous enhancement 
in $h\to \ga\ga$.
Concerning the heavy $\cp$-even Higgs boson, the results are
qualitatively
  similar, but with a smaller possible enhancement of the
  $\ga\ga$ or $WW^{(*)}$ rate, as we show in the right column of
  \reffi{fig:RWW_mssm}. 

\begin{figure}
\centering
\includegraphics[width=0.48\columnwidth]{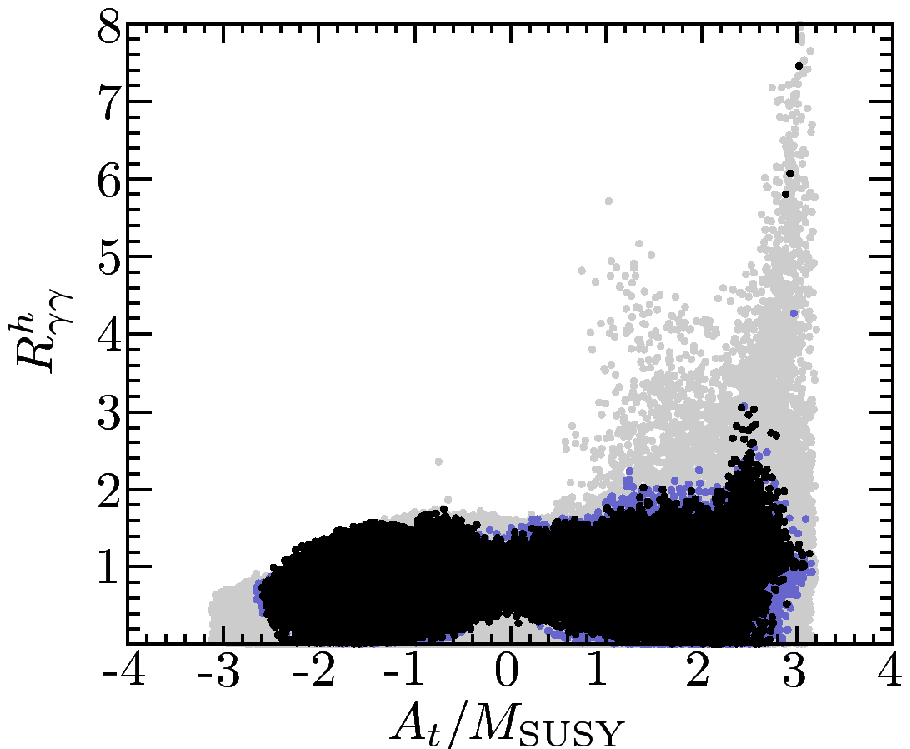}
\includegraphics[width=0.48\columnwidth]{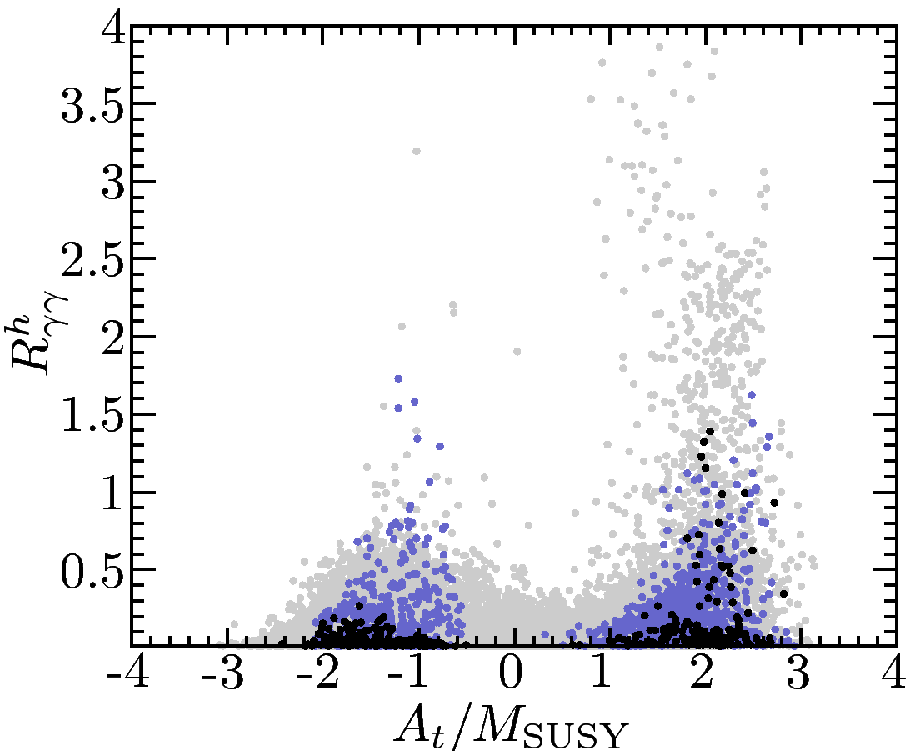}
\caption{Results from the MSSM parameter scan on the ratios
$R^{h}_{\ga\ga}$ (left) and $R^{H}_{\ga\ga}$
  (right) as function of $A_t/M_{\rm SUSY}$. The colour code is as in 
\reffi{fig:rgaga_mssm_br}.}
\label{fig:rgaga_At}
\end{figure}

Finally, in the left (right) plot of \reffi{fig:rgaga_At} we show the
dependence of $R^{h}_{\ga\ga}$ ($R^{H}_{\ga\ga}$) as a function of
$\At/\msusy$, with the 
colour coding as in \reffi{fig:rgaga_mssm_br}. Here it
should be kept in mind that values of $|\At| \gsim \sqrt{6} \, \msusy$
could potentially lead to charge or colour breaking minima~\cite{ccb}. 
For the light $\cp$-even Higgs, as shown in the left plot, the largest
values of $R^h_{\ga\ga}$ are indeed found around $\At \sim \sqrt{6} \, \msusy$. 
However, even applying a stringent cut $|\At| < \sqrt{6} \, \msusy$ 
would still
leave allowed points with $R^h_{\ga\ga} \sim 2$. For the heavy
$\cp$-even Higgs, we see that a stringent cut on $|\At|$ would 
leave nearly all points with $R^H_{\ga\ga} > 1$.


\subsection{$\cp$-even Higgs decays in the NMSSM}

We now turn to the NMSSM and analyse the di-photon decay in this
model. For a light $\cp$-even Higgs boson with a mass below $114.4
\gev$,%
\footnote{
\htg{As before, we neglect here and in the following,} 
the theory uncertainty of the Higgs
boson mass evaluation, which for the light Higgs boson should be
comparable to \htg{or slightly larger than}
the respective uncertainty of the MSSM, i.e.\ roughly at
the level of $2-3 \gev$~\cite{mhiggsAEC}.
}%
~it was pointed out in \cite{Ellwanger:2010nf} that a strong
enhancement of the $\ga\ga$ rate --- up to a factor of seven over the SM
rate --- is possible. In light of the recent results, the main focus of
our analysis will naturally be on higher Higgs masses. For other
approaches regarding an interpretation of
a possible Higgs signal around $\Mh=125\gev$ 
in terms of an NMSSM Higgs boson, see
\cite{Ellwanger2, Mh125NMSSM,Cao:2012fz}.

As before we consider the one-loop induced Higgs decay $h_{1,2}\to
\ga\ga$, but now calculated using the framework described in
\refse{sect:loop}. We perform a scan over the NMSSM parameter space and
evaluate the partial widths and branching ratios for this 
mode. The parameter ranges used for the scan are given in
\refta{tab:hranges}. The remaining parameters are fixed to the benchmark
scenario defined above, see \refeq{eq:MSSMparams}.
It should be noted that the ranges in
\refta{tab:hranges} are not meant to
cover the full NMSSM parameter space. The effects discussed above
that can cause an enhancement of $R^{h,H}_{\ga\ga}$ in the MSSM can be
realised also in the context of the NMSSM. In the present analysis we are
interested in genuine NMSSM effects, which go beyond the MSSM
phenomenology. Such genuine NMSSM effects arise in particular from the 
mixing of the Higgs doublet fields with the Higgs singlet. To be
specific, we consider scenarios that are characterised by large values
of $\MHp$, corresponding to the ``SM$+$singlet'' limit of the NMSSM
discussed above. We furthermore restrict $\mu_{\rm eff}$ and $\tb$
to relatively
small values, while our MSSM scan (cf.~\refta{tab:hrangesMSSM}) had
extended to rather large values of $\mu$ and $\tb$ and had focussed on
the region of relatively low values of $\MA$.
The parameters are chosen such that the mechanisms for enhancing
$R_{\ga\ga}^{h,H}$ realised in the MSSM do not play a role, putting the
emphasis on the genuine NMSSM effects.

\begin{table}[htb!]
\centering
\begin{tabular}{cccl}
\hline
Parameter & Minimum & Maximum \\
\hline
$A_t=A_b=A_\tau$ & $-2\,400$ & $2\,400$ & $\gev$ \\
$\mu_{\rm eff}$ & $150$ & $250$ & $\gev$ \\
\hline
$\MHp$ & $500$ & $1\,000$ & $\gev$ \\
$\tb$ & $2.6$ & $6$ \\
\hline
$\la$ & 0.5 & 0.7\\
$K$ & 0.3 & 0.5\\
$\Aka$ & $-100$ & $-5$ & $\gev$ \\
\hline
\end{tabular}
\caption{Parameter ranges used for the $\cp$-even Higgs decay scan in the
  NMSSM.} 
\label{tab:hranges}
\end{table}

\begin{figure}[htb!]
\centering
\includegraphics[width=0.48\columnwidth]{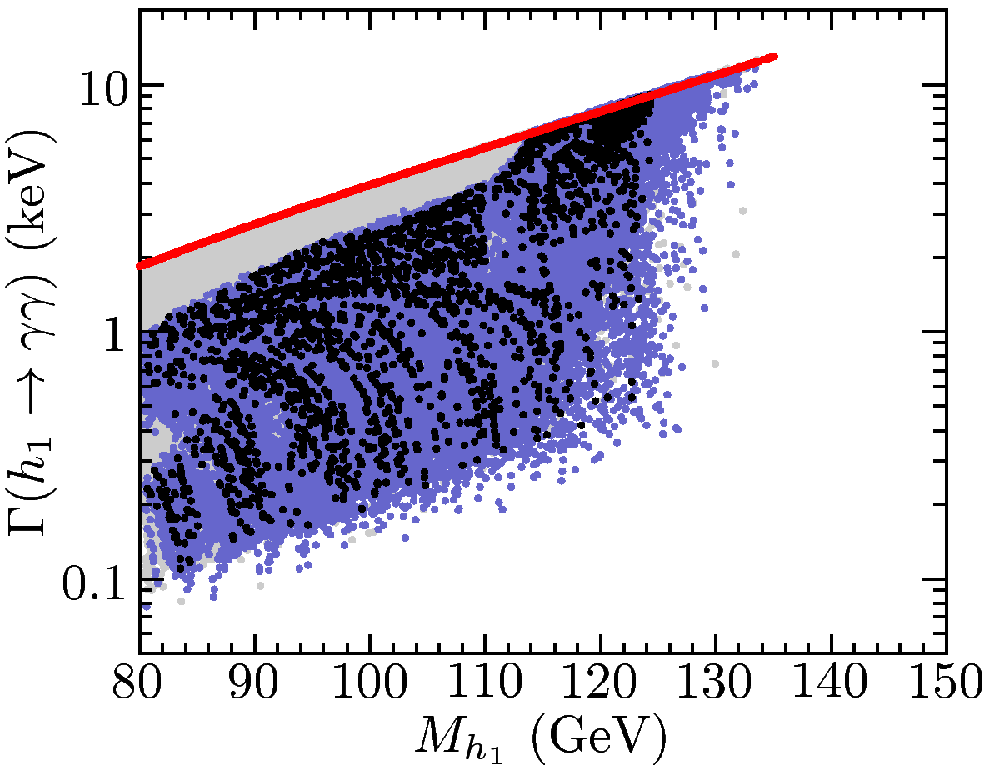}
\includegraphics[width=0.48\columnwidth]{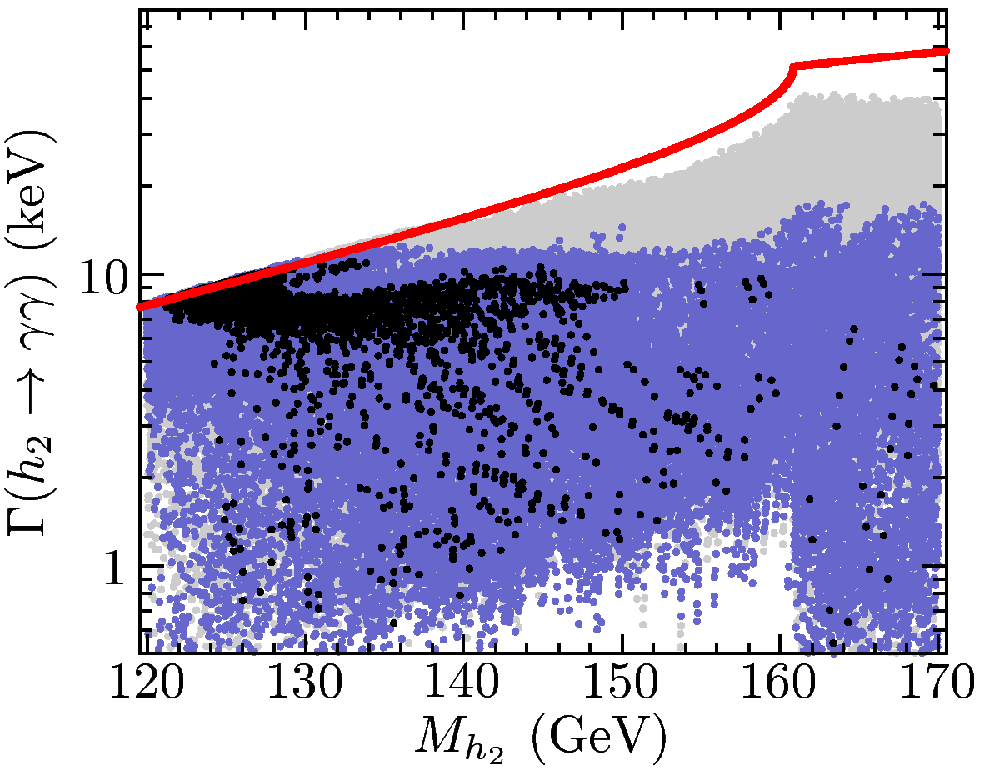}\\
\includegraphics[width=0.48\columnwidth]{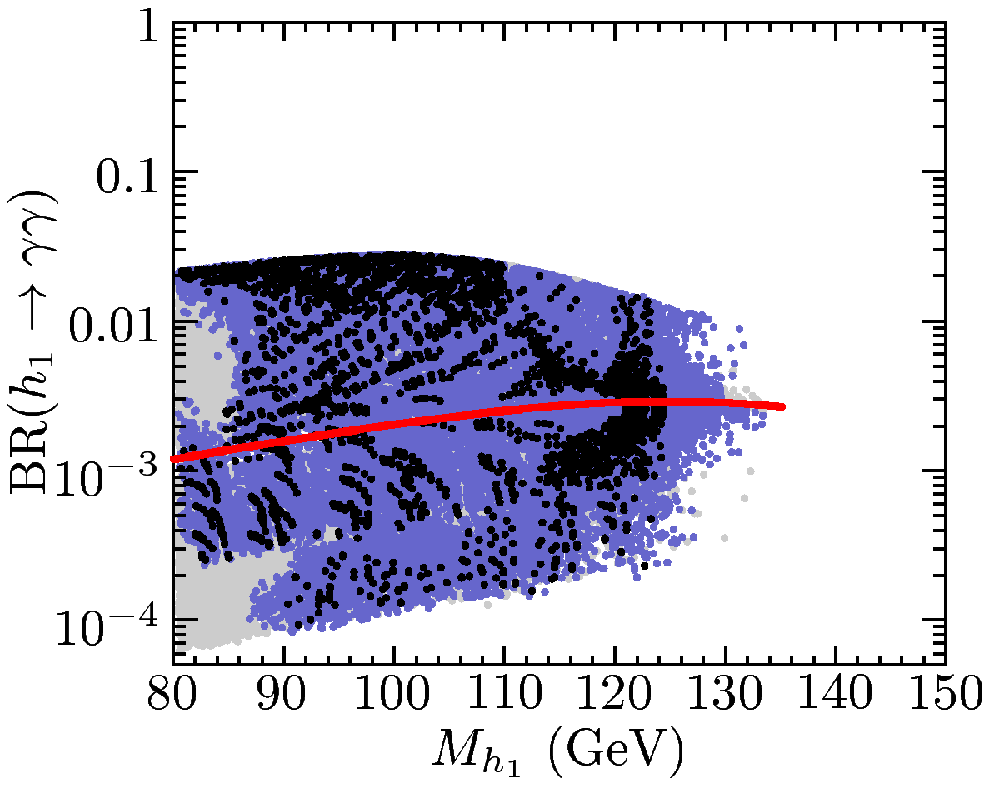}
\includegraphics[width=0.48\columnwidth]{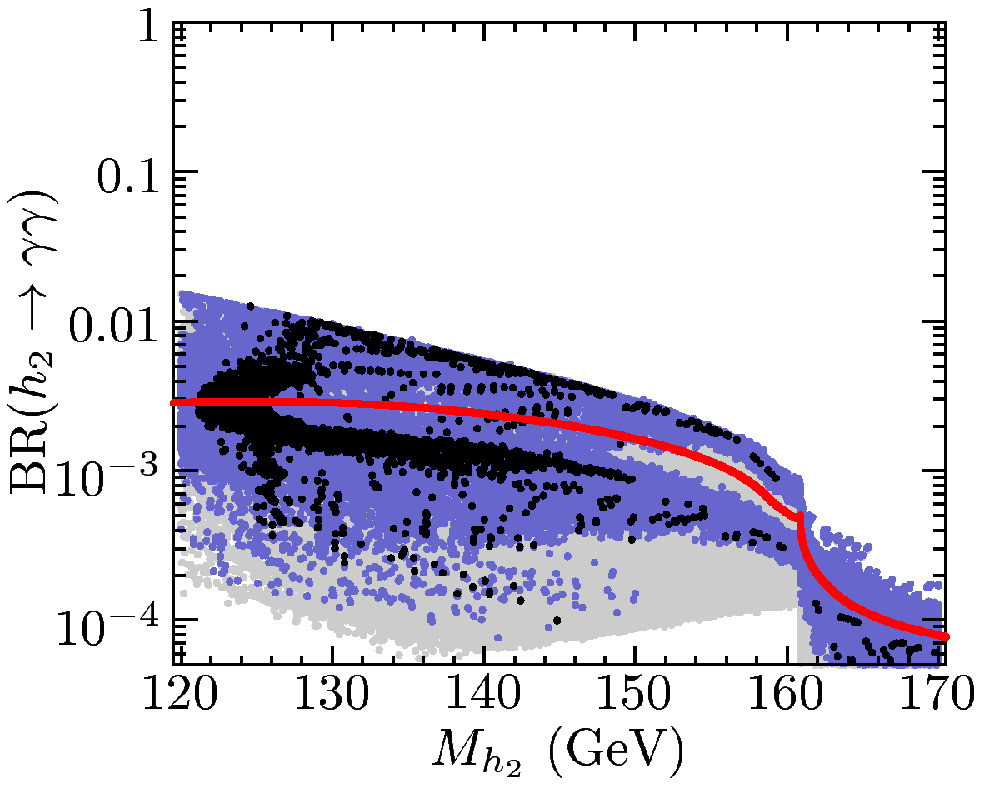}
\caption{Results from the NMSSM parameter scan (see text)
for the partial widths
  $\Ga(\Hi\to\ga\ga)$ and the corresponding branching ratios of
  $\He$ (left) and $\Hz$ (right).  
All points in the figure fulfil the ``theoretical'' constraints defined
in \refse{sect:constraints}. In addition, the blue (dark) points
satisfy direct Higgs search limits from colliders
\htg{(from {\tt HiggsBounds}~3.6.1, i.e.\ including \lhcee)}, 
while the black
points are in agreement with all theoretical and experimental
constraints. The solid (red) curve shows the respective quantities
evaluated in the SM.} 
\label{fig:gaga}
\end{figure}

The results for $\Hez\to \ga\ga$ are shown in
\reffi{fig:gaga} ($h_3$ is always heavy and plays no role in our analysis). 
The colour coding is the same as in \reffi{fig:rgaga_mssm_br},
i.e.\ all displayed points satisfy the
``theoretical'' constraints of \refse{sect:constraints} (grey). Points 
which in addition fulfil the direct Higgs exclusion
limits from colliders \htg{(from {\tt HiggsBounds}~3.6.1,
i.e.\ including \lhcee)} are drawn in blue, and points which
satisfy all the constraints, in particular also those from 
$\De a_\mu$ and flavour physics, are shown in black. The red curve shows
the corresponding SM result, obtained by setting $\MHSM = \mHez$,
respectively.

We choose to study $\mHe$ in a range from $80 \gev$ up to its maximum
around $135 \gev$. Allowed points with $\mHe < 80 \gev$ are also found
in the scan, but the large singlet component of these very light Higgs
bosons gives rise to a quite different phenomenology, which we do
not investigate in detail here.
For masses close to $140 \gev$, the number of
allowed points is seen to decrease, which illustrates that only quite
specific choices of the input parameters give $\mHe$ close to the
maximum. This, as well as other features with local under- (over-) 
density of points in certain regions, can simply be viewed as
sampling artefacts, i.e.\ the point density has no physical meaning.
For $\Hz$ we study the mass interval $120 \gev < \mHz < 170 \gev$, 
which means there is an overlap with the region considered for
$\mHe$. To go even higher in
$\mHz$ is not particularly interesting for our purposes, 
since when the two-body decay
$\Hz \to WW^{(*)}$ is open the loop-induced $\Hz\to \ga\ga$ decay becomes
suppressed (as is also clearly visible in the figure).  

\reffi{fig:gaga} shows that $\Ga(\Hi\to \ga\ga)$ is always
smaller than (or at most equal to) its SM value for the points in our scan. 
This means in particular that our scan, for which we have fixed
the slepton masses to large values (see \refeq{eq:MSSMparams}), does not
contain points with light staus (the contribution of light staus was
discussed in the MSSM context above).
For $\mHz \gsim 140 \gev$, the partial width does not
reach the full SM value, which shows that this
mass region is not accessible for a fully SM-like $\Hz$. Taking into
account the collider 
constraints, we also see that, \htg{as in the MSSM, }
a SM-like Higgs boson 
with $\mHe < 114.4 \gev$ is excluded as a consequence of the LEP
limits.
Despite the smaller NMSSM width for
$\Ga(\Hi\to \ga\ga)$ compared to the SM, \reffi{fig:gaga}
shows that an enhancement of the branching ratio with up to an order of
magnitude over the SM is possible. 
The results are similar for $h_1$ and $\Hz$ in the overlapping mass
region.

\begin{figure}
\centering
\includegraphics[width=0.48\columnwidth]{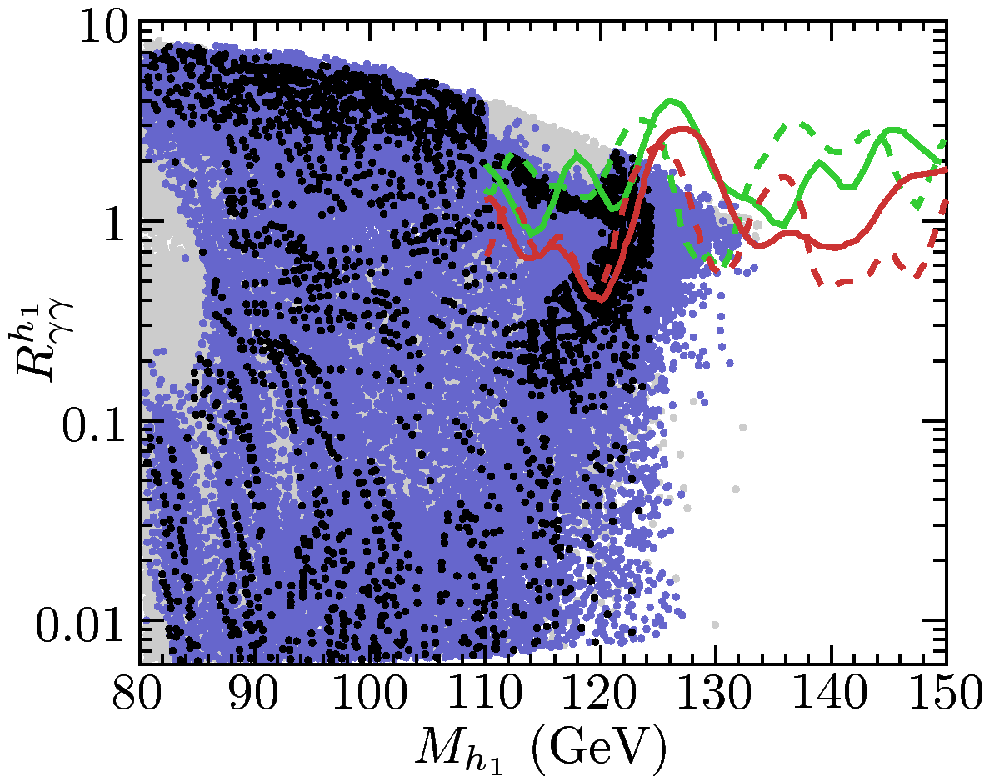}
\includegraphics[width=0.48\columnwidth]{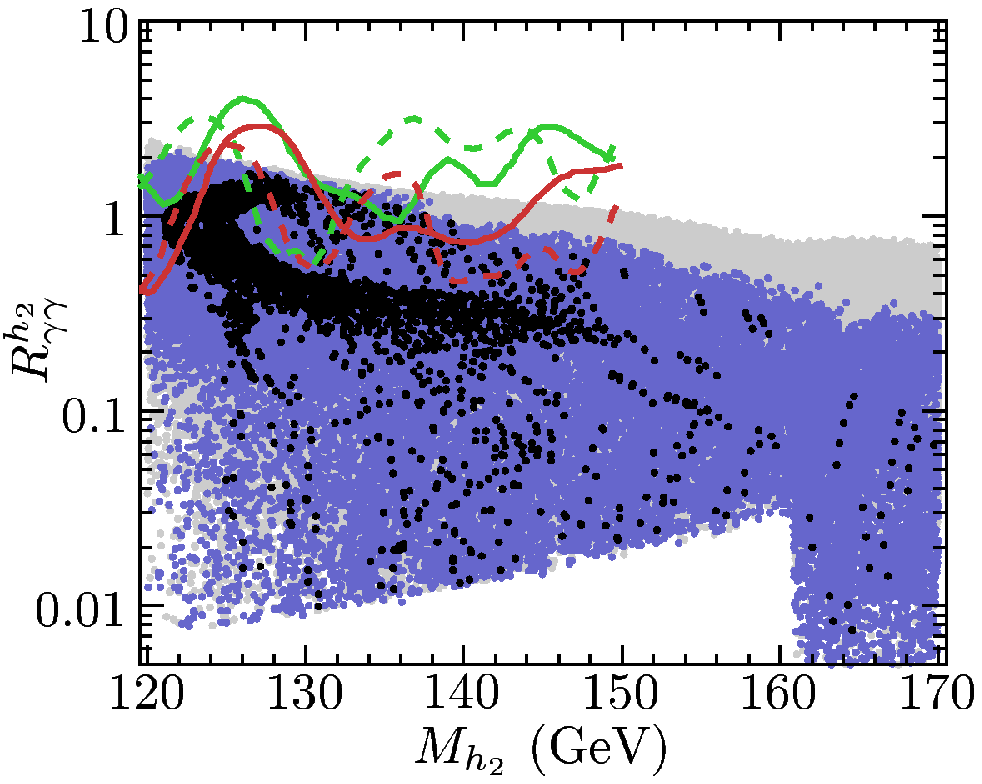}\\
\includegraphics[width=0.48\columnwidth]{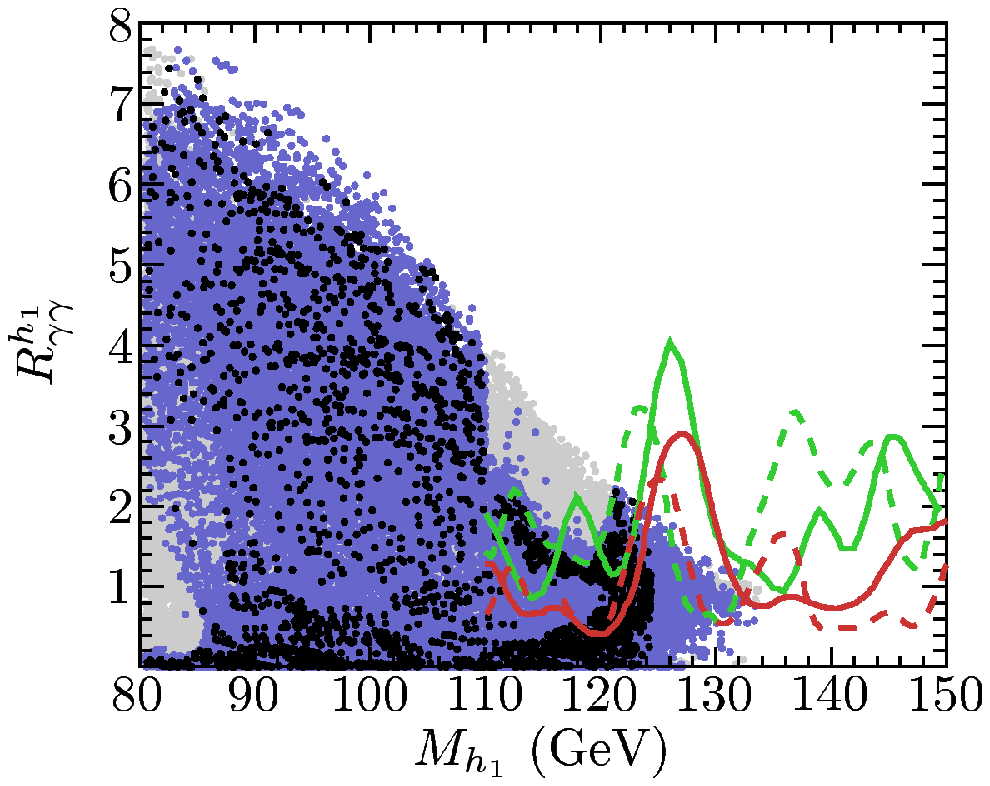}
\includegraphics[width=0.48\columnwidth]{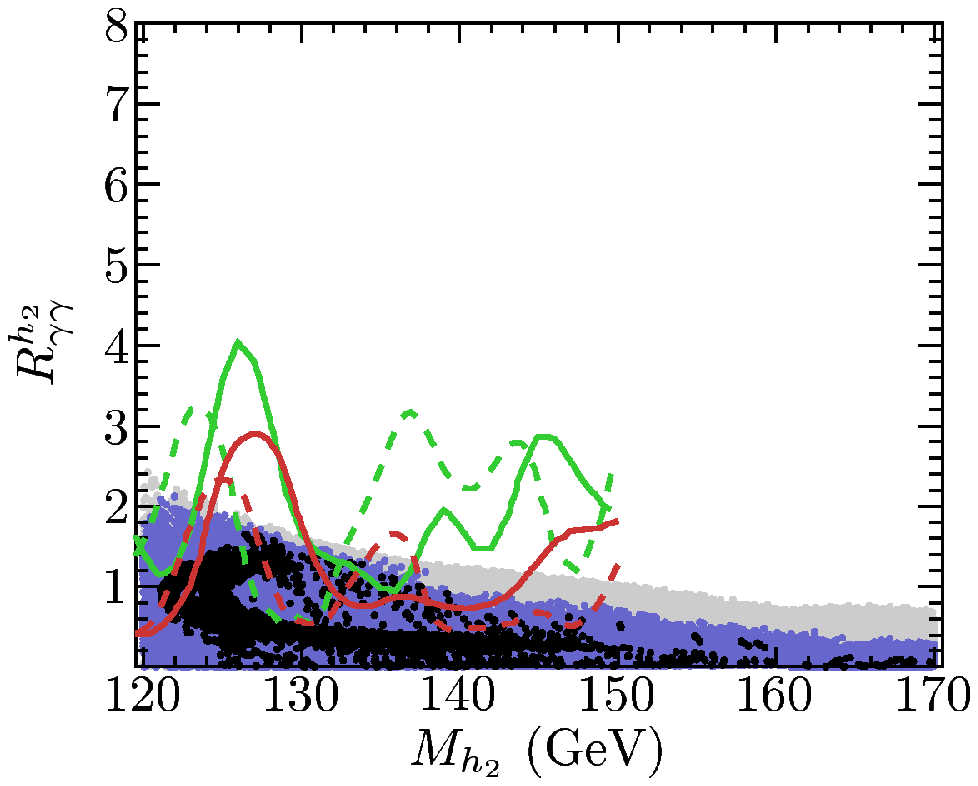}
\caption{Results from the NMSSM parameter scan on the ratio
  $R^{\Hi}_{\ga\ga}$ for the two lightest Higgs bosons $h_1$
  (left column) and $\Hz$ (right column). 
  The plots are displayed both on a logarithmic scale
(upper row) and on a linear scale (lower row).
  The colour coding for the scan points is the same as in
  \reffi{fig:gaga}. The green lines show 
  exclusion limits on this channel at $95\%$ CL from \htg{\lhcee\ data from}
 ATLAS~\cite{ATLAS_gaga_Dec11} (solid) and from
  CMS~\cite{CMS_gaga_Dec11} (dashed).
  \htg{The red lines are the new limits from ATLAS (solid) and CMS (dashed)
    taken from~\cite{discovery}.}
}
\label{fig:rgaga_mh1}
\end{figure}

As in the case of the MSSM we now analyse $R_{\ga\ga}^{\Hi}$. 
The total
widths appearing in \refeq{eq:Rgaga} are calculated in an
approximate way according to 
\begin{align}
\Ga_{\mathrm{tot}}(\Hi)=\frac{1}{\mHi}
\mathrm{Im}\Bigl[\,\Sigma_{\Hi}(\mHi^2)\Bigr]+\Ga(\Hi\to WW^{(*)})
+\Ga(\Hi\to \ga\ga)+\Ga(\Hi\to gg), 
\end{align}
where $\Sigma_{\Hi\Hi}$ denotes the one loop self energy of $\Hi$. The
inclusion of the off-shell decays, as well as the loop-induced
processes, in the total width is essential for a 
realistic prediction.

The results for $R^{\He}_{\ga\ga}$ and
$R^{\Hz}_{\ga\ga}$ from the scan over the NMSSM parameter space
are shown in \reffi{fig:rgaga_mh1}. As before, we show the plots
both on a logarithmic and a linear scale, \htg{and include the latest
  results on this channel taken from~\cite{discovery}}. 
Looking first at $\He$, the figure shows that a sizable enhancement over
the SM rate is possible
over the whole mass range from $\mHe=80\gev$ to $\mHe=130 \gev$. For the
range of Higgs masses 
below the SM limit, $\mHe < 114.4 \gev$, points with a significant
enhancement $R^{\He}_{\ga\ga} \gsim 7$ are observed, in accordance
with the results of \cite{Ellwanger:2010nf} (see also
\cite{Cao:2011pg}).

Turning to $\Hz$, the results for $R_{\ga\ga}^{\Hz}$ are similar to those for
$R^{\He}_{\ga\ga}$ in the common mass range; the observed maximal
enhancement is $R^{\Hz}_{\ga\ga} \gsim 2$ for $\mHz$ in the range from
$120\gev$ to $125 \gev$.  
A smaller enhancement over the SM 
is possible for all
$\mHz < 145 \gev$. As $\mHz$ approaches $160 \gev$, where the on-shell
decay $\Hz\to WW^{(*)}$ opens, the rate drops to
$R^{\Hz}_{\ga\ga}<1$.

Comparing the results for $R_{\ga\ga}^{\Hi}$ to the limits
\htg{most recent limits from ATLAS (solid red) and CMS (dashed 
red)~\cite{discovery} (where the \lhcee\ limits are shown in green)}, 
it is clearly visible that the
NMSSM (similarly to the MSSM) can produce points with a large
suppression of $R_{\ga\ga}$. 
Concerning the more interesting case $R^{\Hi}_{\ga\ga} > 1$, we see
that the NMSSM can produce an enhancement compatible with an excess
over the SM rate for Higgs production in the 
mass region around $125 \gev$. 
The results of \reffi{fig:rgaga_mh1} show that such 
\htg{observed excess over the SM rate for Higgs production in the
  $\ga\ga$ channel is well compatible 
with both $\He$ or $\Hz$ production in the NMSSM.}

\begin{figure}[tb!]
\centering
\includegraphics[width=0.32\columnwidth]{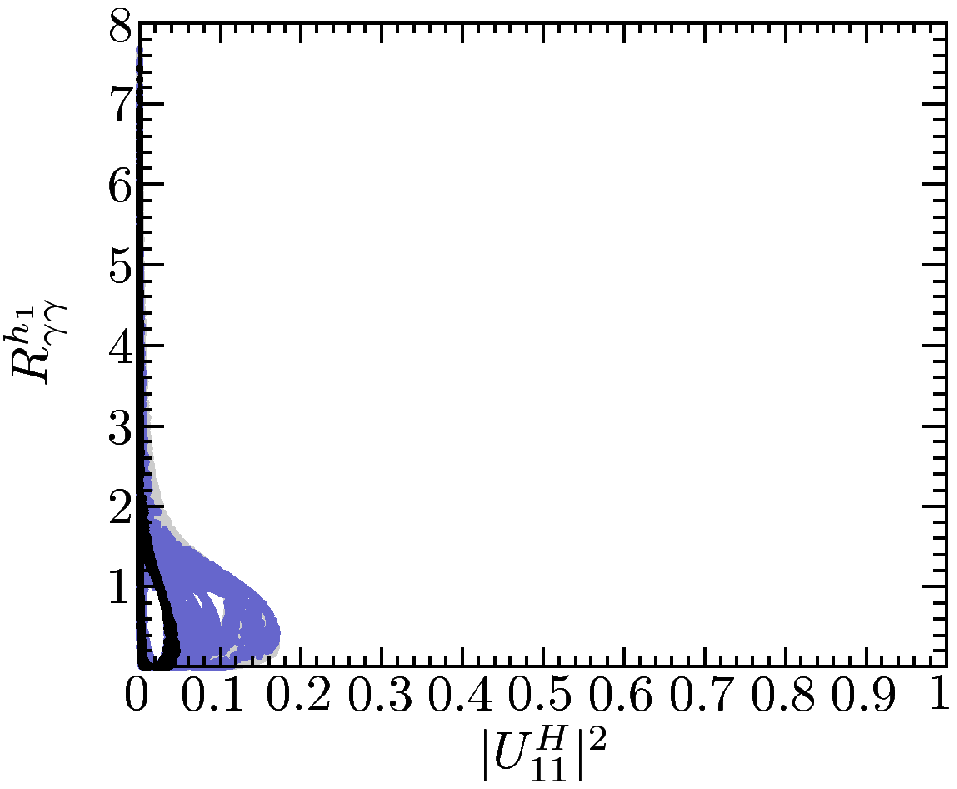}
\includegraphics[width=0.32\columnwidth]{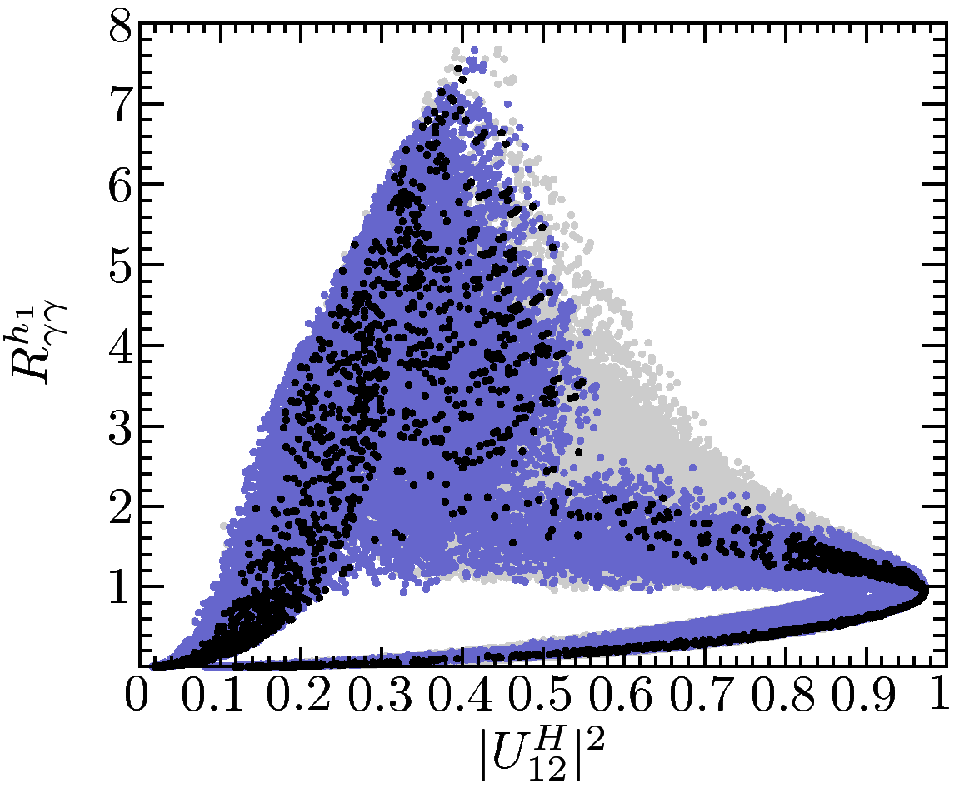}
\includegraphics[width=0.32\columnwidth]{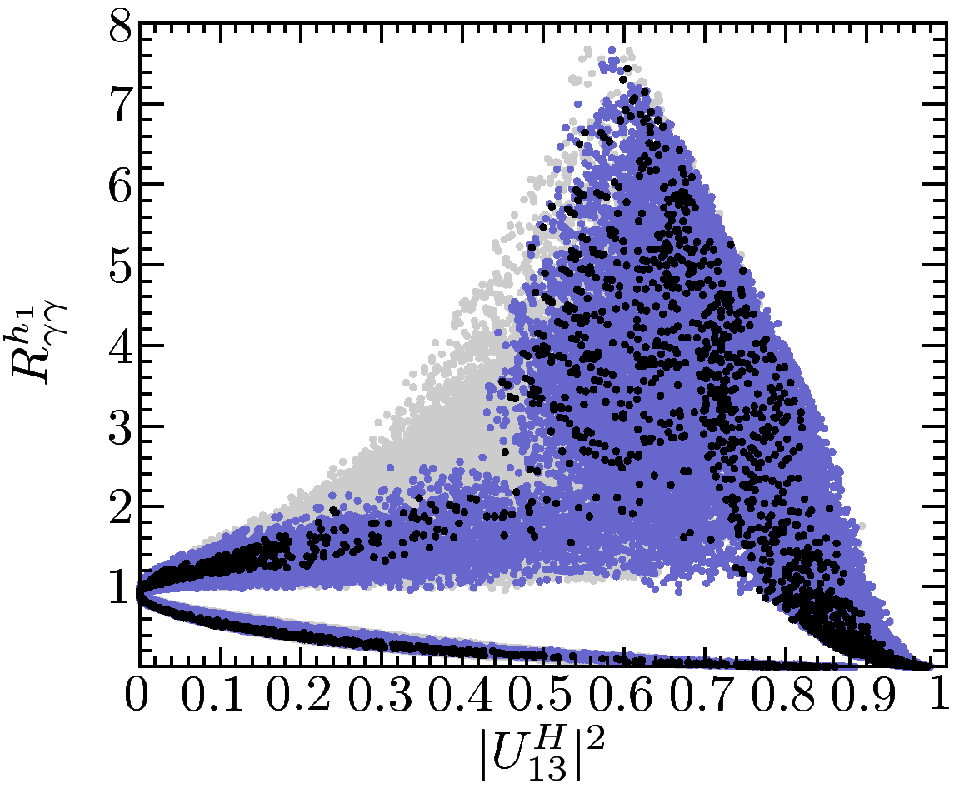}\\
\includegraphics[width=0.32\columnwidth]{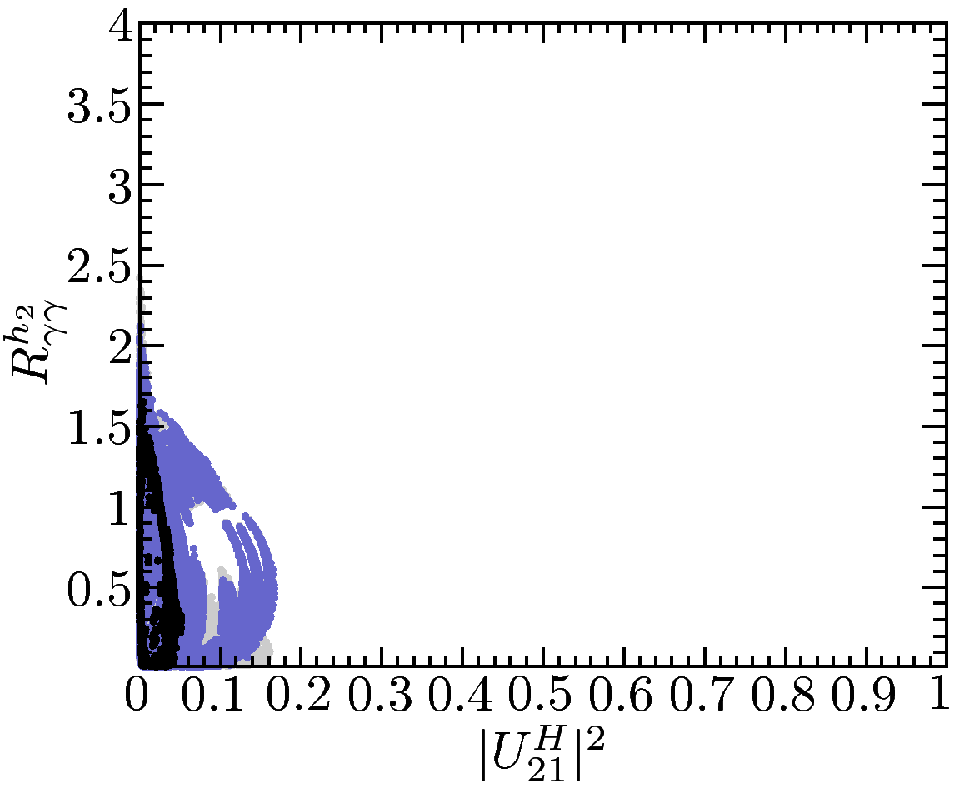}
\includegraphics[width=0.32\columnwidth]{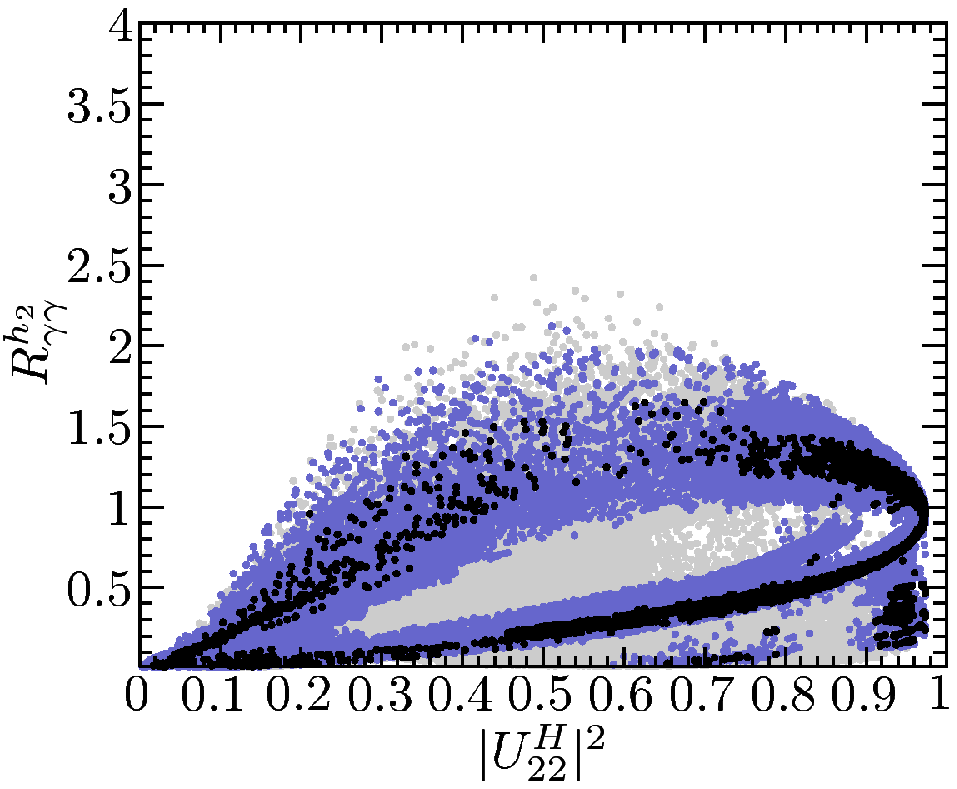}
\includegraphics[width=0.32\columnwidth]{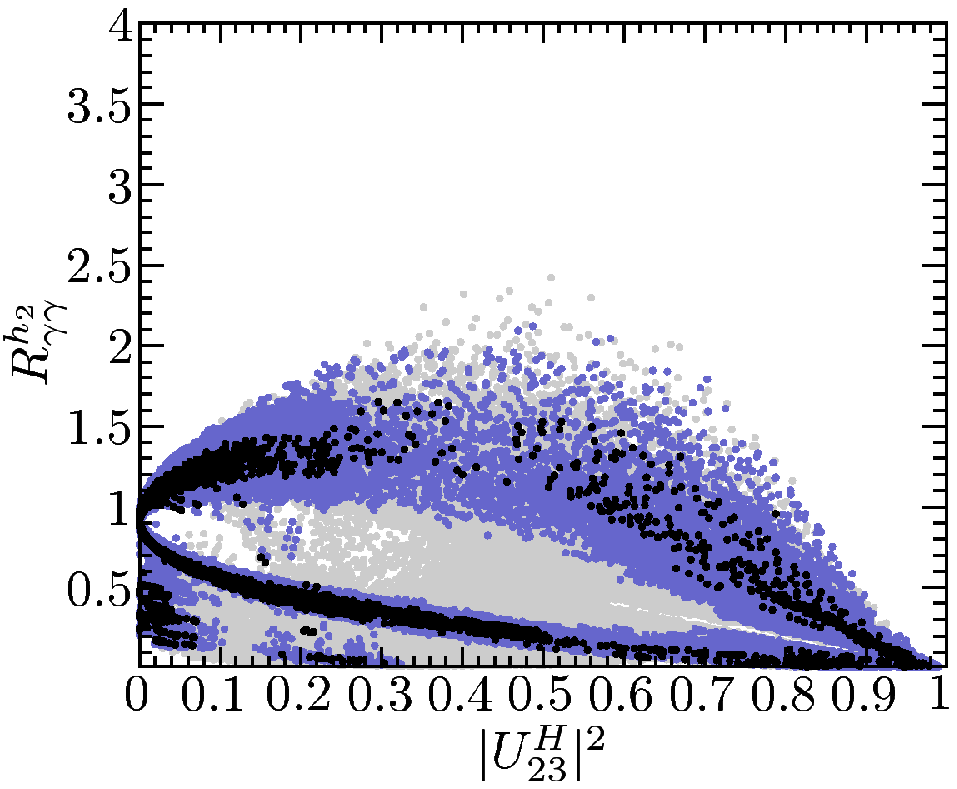}
\caption{Results from the NMSSM parameter scan on
  $R^{\Hi}_{\ga\ga}$ and the elements $\Ueven_{ij}$ of the $\cp$-even
  Higgs mixing matrix  for $\He$ (top) and $\Hz$ (bottom). The colour
  coding is the same as in \reffi{fig:gaga}.}
\label{fig:rgaga_zh}
\end{figure}

In order to identify the conditions under which a significant NMSSM enhancement
of $R^{\Hi}_{\ga\ga}$ is possible --- as explained above, this is a
genuine NMSSM effect that goes beyond the mechanisms discussed above for 
the MSSM --- we show in
\reffi{fig:rgaga_zh} the dependence of $R^{\Hi}_{\ga\ga}$ on the composition of
$\Hi$ as defined in \refeq{eq:RotateToMassES}. 
\reffi{fig:rgaga_zh} shows that an important
requirement for $R^{\Hi}_{\ga\ga} > 1$ is that $\Ueven_{i1} \simeq 0$, which
means that the corresponding Higgs mass eigenstate lacks a $H_d$
component. In the limit $\Ueven_{i1} \to 0$ (corresponding to $\aeff \to
0$ in the MSSM) the otherwise dominant decay channels $\Hi \to b\bar b$
and $\Hi \to \tau\tau$ vanish (in the effective coupling
approximation), 
thereby increasing $\br(\Hi\to\ga\ga)$. Since $\Ueven$ is unitary, the general
sum rule 
\begin{align}
\sum_j|\Ueven_{ij}|^2=1
\end{align}
implies that points with $\Ueven_{i1}=0$ must have
$|\Ueven_{i2}|^2+|\Ueven_{i3}|^2=1$. From \reffi{fig:rgaga_zh} it can be
seen that a configuration that maximises $R_{\ga\ga}^{\Hi}$ would be
\begin{align*}
|\Ueven_{i1}|^2 = 0\,,\qquad 
|\Ueven_{i2}|^2 \simeq 0.4\,, 
\qquad |\Ueven_{i3}|^2\simeq 0.6~.
\end{align*}
Unlike the case of Higgs doublet mixing resulting in a small $\aeff$ in
the MSSM (which requires a low value for $M_A$ and a high $\mu$),
in the NMSSM this feature is caused by a sizable singlet component
of $\Hi$. The observed $R_{\ga\ga}^{\Hi}$ enhancement is therefore a
genuine feature of the NMSSM which is still present even in the 
SM+singlet limit. On the other hand, in the MSSM limit (where
$|\Ueven_{i3}|^2=0$) points from our scan show only very small
$R_{\ga\ga}^{\Hi}$ enhancements. This is a consequence of the fact
that we have restricted our scan in the NMSSM to large $\MHp$, large
slepton and squark masses, as well as to relatively small values of 
$\tb$ and $\mu_{\rm eff}$, which corresponds to a parameter region in
the MSSM that is complementarity to the one used for our MSSM scan.

\begin{figure}[t!]
\centering
\includegraphics[width=0.48\columnwidth]{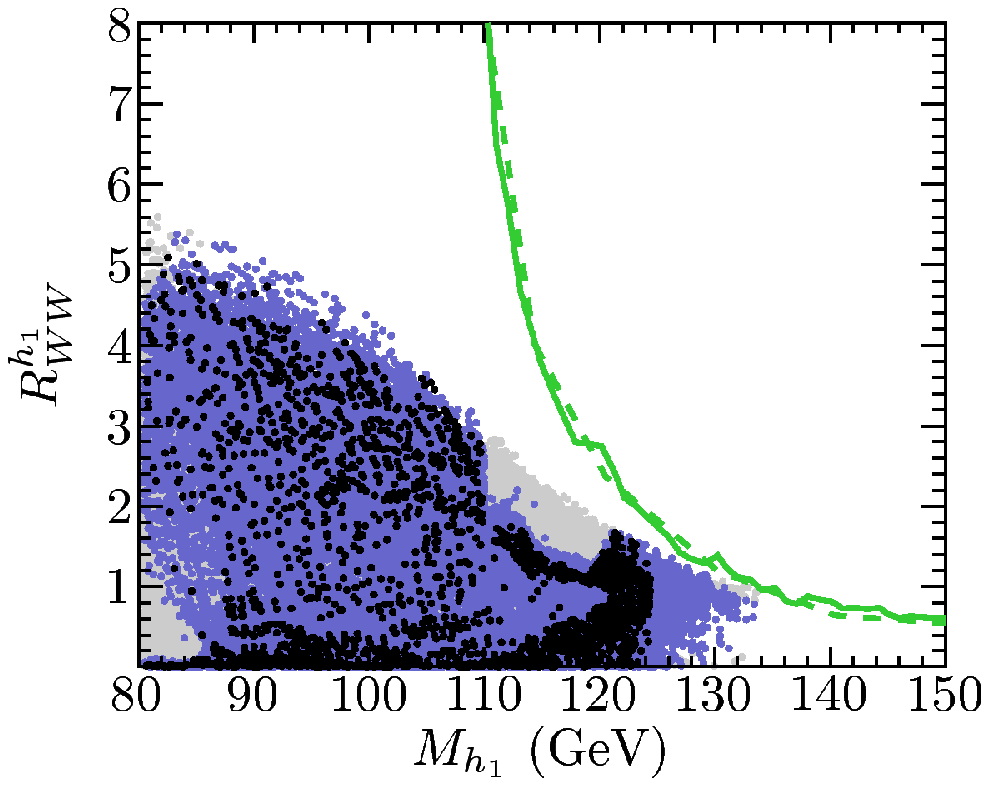}
\includegraphics[width=0.48\columnwidth]{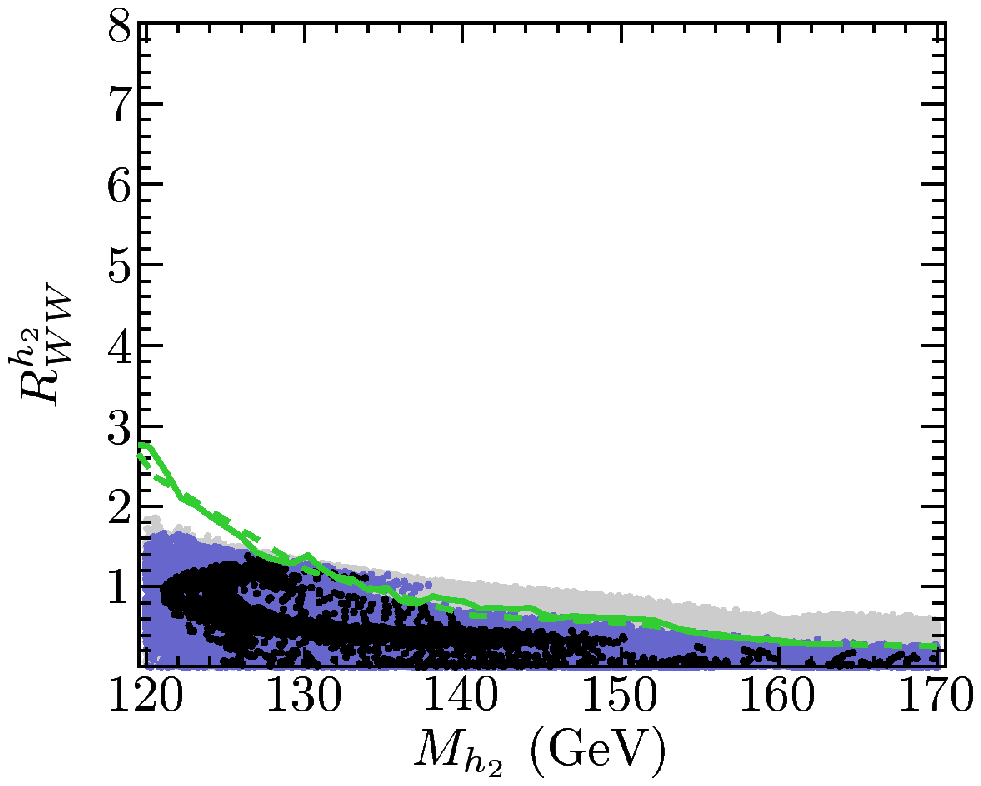}\\
\includegraphics[width=0.48\columnwidth]{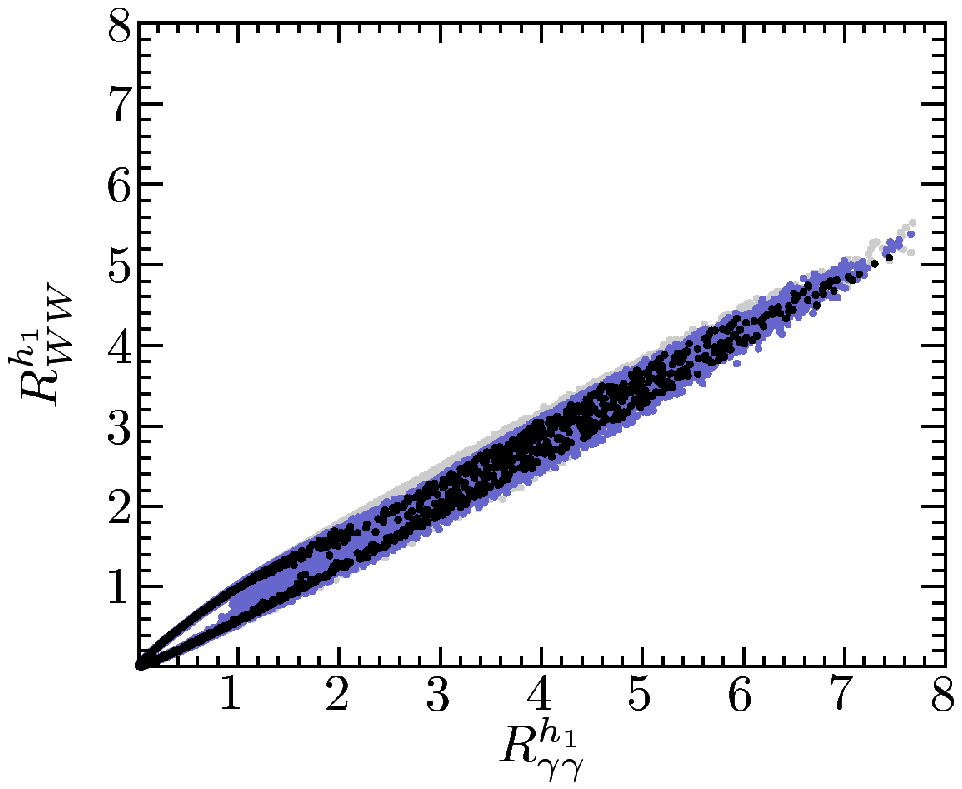}
\includegraphics[width=0.48\columnwidth]{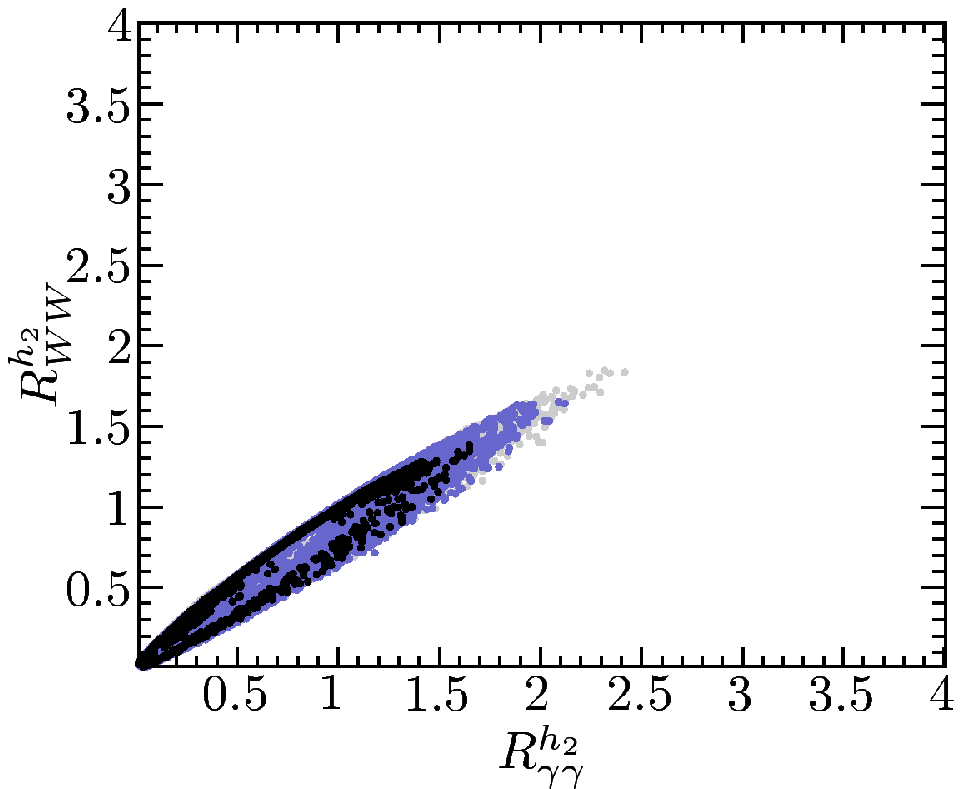}
\caption{Results from the NMSSM parameter scan on the ratios $R^{h_i}_{WW}$
($i = 1,2$) for the NMSSM Higgs bosons $\He$ (left column) and
  $\Hz$ (right column). The plots in the upper row show $R^{h_i}_{WW}$
as a function of the respective Higgs mass. The $95\%$ CL
exclusion limits for the $WW^{(*)}$ channel from \htg{\lhcee\ data from}
ATLAS~\cite{ATLAS_WW_Dec11}
(solid line) and CMS~\cite{CMS_WW_Dec11} (dashed line) are also shown.
The plots in the lower row show the correlation of
  $R^{h_i}_{WW}$ and $R^{h_i}_{\ga\ga}$. The colour
  coding is the same as in \reffi{fig:gaga}.
}  
\label{fig:RWW}
\end{figure}

As for the MSSM case, also
in the NMSSM the decay $\Hi\to \ga\ga$ is usually dominated by
contributions from loops containing $W$ bosons, and we expect a
corresponding correlation of $R^{\Hi}_{\ga\ga}$ with
the (off-shell) decays $\Hi\to WW^{(*)}$ and $\Hi\to ZZ^{(*)}$.
This is studied in \reffi{fig:RWW}, where we
give results for the tree-level decays $\Hez\to WW^{(*)}$ using the same
colour coding as above. As expected, a sizable enhancement is
possible for $R^{\Hi}_{WW}$, in particular for $\He$, and a strong
positive correlation between $R^{\Hi}_{WW}$ and
$R^{\Hi}_{\ga\ga}$ is visible. 
The possibility of a simultaneous enhancement of these two modes can
again be understood as an effect of the large suppression of the main
fermionic coupling $\Hi b\bar{b}$, which leads to an enhancement of the
respective branching ratios. As in the case of the MSSM, despite the
positive correlation between $R^{\Hi}_{WW}$ and $R^{\Hi}_{\ga\ga}$ it is
nevertheless possible to have a both a slight enhancement of 
$R^{\Hi}_{\ga\ga}$ and a slight suppression of $R^{\Hi}_{WW}$.

In \reffi{fig:RWW} we also display the ATLAS~\cite{ATLAS_WW_Dec11} and
CMS~\cite{CMS_WW_Dec11} limits on the $H \to WW$ channel \htg{based on
  \lhcee. For the same reasons as in the MSSM case we do not include the
  new results from~\cite{discovery}. The \lhcee\ } limits 
exclude a SM-like Higgs for $\MH \gtrsim 130 \gev$
\htg{(which would not change by the inclusion of the data
from~\cite{discovery})}. The points excluded 
by this mode 
can be seen as a grey strip for
$\mHz\gtrsim 130\gev$ in this figure and also in \reffi{fig:rgaga_mh1}. 
For the few points above the green lines which are not excluded by $H
\to WW$, another channel has a higher expected sensitivity. 
For $\mHi<130 \gev$ the \htg{existing} limits on $\Hi\to WW$ do not pose any
constraint on the NMSSM, but an improved exclusion of the $WW$ mode at
lower Higgs masses could have severe implications for the scenarios we
consider. The resulting limits on $R^{\Hi}_{\ga\ga}$ \htg{are}
competing with those from direct searches in this channel, 
\htg{but appear somewhat relaxed in view of the excess over the SM
prediction in the $\ga\ga$ channel as reported in~\cite{discovery}.
The combination of those two}
channels offers an excellent opportunity for cross-checks and model testing.


\newcommand{\phiA}{\phi_A}
\newcommand{\mphiA}{m_{\phiA}}

\subsection{$\cp$-odd Higgs decays in the NMSSM and the MSSM}

For completeness,
we have also investigated decays of $\cp$-odd Higgs
bosons in the MSSM and in the NMSSM.  
In the following we denote a generic
$\cp$-odd Higgs boson as $\phiA$. 
As a consequence of gauge invariance, the $\ga \phiA Z$ vertex vanishes
in both models at the 
tree-level. The vertices $VV\phiA$ ($V=\ga, g, Z, W$) vanish 
by virtue of $\cp$-invariance~\cite{Keller:1986nq}. 
Accordingly, all these vertices only appear at the one-loop level.

The MSSM contribution to $A\to \ga\ga, gg$ has
been evaluated by many groups (see \cite{MSSMHiggsRev} for a
comprehensive list of references).
The  MSSM prediction for $A\to WW $ was first studied in 
\citeres{Gunion:1991cw,Chankowski:1992es}, however only heavy SM
fermions were included. In \citere{Chankowski:1992es}, chargino and neutralino
contributions to $A\to WW $ were included together with the SM fermions.
In the MSSM
the BR's of $A\to \ga\ga$ and $A\to WW $ are of \order{10^{-4}} and
\order{10^{-3}}, respectively,
for small $\tb$ and decrease for large $\tb$. 
Consequently, for a production cross section similar to a SM Higgs boson with
the same mass, $R^A_{\ga\ga}$ can be order of \order{10^{-2}}. The
prospects for the observation of those
modes at the LHC are not very encouraging.

In order to assess the NMSSM potential for the observation of the $\cp$-odd
Higgs bosons in these loop-induced decay channels, a random scan over
the six-dimensional parameter space of the Higgs 
sector has been performed (see \refta{tab:hranges}).
In this scan the lightest $\cp$-odd Higgs boson has a mass ranging
from about $80$ to $250 \gev$. Due to the mixing of doublet and singlet
states, \refeq{eq:RotMat}, the NMSSM  $\cp$-odd Higgs bosons can have couplings
to fermions that are significantly smaller than in the MSSM. 
Contrary to our analysis in the previous section where we compared to
the SM, we now compare the NMSSM results for the light $\cp$-odd Higgs
boson $\Ae$ to the results for the 
$\cp$-odd Higgs boson in the MSSM (with $\mAe = \MA$). 
Both in the MSSM and the NMSSM, we compute the total decay widths of
$\cp$-odd Higgs according to
\begin{align}
\Ga_{\mathrm{tot}}(\phiA)=\frac{1}{\mphiA}
\mathrm{Im}\Bigl[\,\Sigma_{\phiA}(\mphiA^2)\Bigr]
 +\sum_{V=\ga, g, Z, W} \Ga(\phiA \to VV).
\end{align}
In order to compare the NMSSM to the MSSM we define the
ratio of di-photon cross-section normalised to the MSSM rate as follows 
\begin{align}
\hat R_{\ga\ga}^{\Ai} = \frac{
  \KKL \si(gg \to \Ai) \times \br(\Ai \to \ga\ga) \KKR_{\rm NMSSM}}
 {\KKL \si(gg \to A) \times \br(A \to \ga\ga) \KKR_{\rm MSSM}}~.
\label{Rhat} 
\end{align}

In the case where $\la, \ka \to 0$ (with $\ka/\la$ held constant), 
which corresponds to the MSSM limit (see above),
the interactions of the singlet field with the MSSM Higgs doublets
vanish, i.e.\ the singlet field decouples. The decoupled singlet field
itself has no gauge couplings. In this MSSM limit, one finds that
one of the NMSSM $\cp$-odd Higgs bosons, $\Ak$ (which does not
necessarily coincide with $\Ae$)
behaves like the MSSM $A$~boson. The MSSM phenomenology is fully 
retrieved in the Higgs sector, the total decay width 
of the $\cp$-odd Higgs in the NMSSM approaches the MSSM value, 
and we find $\hat R_{\ga\ga}^{\Ak} \sim 1$. 
However, even in this limit, in some cases the
distinction of the two models might still be possible via the neutralino
sector~\cite{MoortgatPick:2005vs}.

Far from the MSSM limit, the coupling
$\Ak f \bar{f}$, which is proportional to the doublet-singlet mixing,
gets suppressed and the NMSSM phenomenology can differ from the MSSM. 
In our scan we find an enhancement of the $\br(\Ae \to \ga\ga)$
of up to 100 over the MSSM. However, also the $\Ae$ production rate gets
suppressed in this case, leading to (at most) $\hat R^{\Ae}_{\ga\ga} \sim 2$. 
Consequently, as for the MSSM, also $\Ae$ is expected to remain 
unobservable in this
channel at the LHC (and we do not show the plots for this channel).


\newpage
\section{Conclusions}
\label{sect:conclusions}

The discovery of a new particle with properties that are compatible with
a Higgs boson has started a new era, with the goals to reveal whether
the new particle is indeed a Higgs boson and whether its detailed
phenomenology is consistent with the SM predictions or whether there is
evidence for physics beyond the SM. The observations reported both by
ATLAS and CMS that the new state has a signal strength in the $\ga\ga$ 
channel that is considerably higher than the one expected in the SM 
may be a first hint into this direction, although the statistical
significance of this deviation from the SM prediction is not
sufficien at present to draw a definite conclusion. 
In view of the possibility of an enhanced rate in the $\ga\ga$ channel,
it is nevertheless interesting to investigate the predictions of
possible alternatives to the SM in this context. We have carried out
such an analysis in the present paper, comparing the predictions for
Higgs boson production in gluon fusion, the main production channel at
the LHC, and its subsequent decay into two
photons in the SM, the MSSM, and the NMSSM. We have furthermore analysed
the $WW^*$ channel, which 
\htg{is strongly correlated with the $\ga\ga$ mode.}

While for the predictions in the SM and the MSSM we have used the
well-known code {\tt FeynHiggs}, for the predictions in the NMSSM we
have developed a new framework consisting in particular of an
appropriate model file for the program {\tt FeynArts} and the input
on masses, mixing angles, etc.\ needed for the numerical evaluation.
In this set-up we can make use of the highly automated programs 
{\tt FormCalc} and {\tt LoopTools} for the evaluation of the relevant
observables. In the present implementation \NMSSMTools\ is used for
higher-order corrections to the masses of the external Higgs bosons.
Numerous tests have been performed to verify the implementation; among
other things we evaluated more than $150$ loop-induced processes in the
NMSSM and checked the results for UV-finiteness. 

We have presented results for Higgs-boson production in
gluon fusion and its decays into $\ga\ga$ and $WW^*$ within the MSSM and
the NMSSM, normalised to the SM prediction. 
We have analysed in detail possible mechanisms for the enhancement (but
also the suppression) of those channels in both models. In this context
we have investigated in particular whether an enhancement of the 
$\ga\ga$ rate for a Higgs mass of about $125\gev$ is compatible with
existing limits on the parameter space arising from theoretical
constraints as well as from the limits from direct searches for
supersymmetric particles, from the Higgs searches at LEP, the Tevatron
and the LHC (where we have incorporated the limits based on the data
\htr{presented} in 2011), from electroweak precision observables and from
flavour physics. Performing parameter scans in both models, we 
have then confronted the points passing all the above constraints with
the latest results of the Higgs searches in the $\ga\ga$ channel that
have just been announced by ATLAS and CMS.
\htg{We have found that an enhanced rate of Higgs
production and decay to two photons can easily be realised in both models, the MSSM as
well as in the NMSSM. This holds not only for the lightest $\cp$-even
Higgs boson in the models, but also for the second lightest $\cp$-even
Higgs boson in both the MSSM and the NMSSM. In this latter
interpretation in both models the lightest
$\cp$-even Higgs boson possesses a strongly suppressed coupling to gauge
bosons and escapes all existing direct searches.

Within the MSSM we have analysed the mechanisms that can lead to such a
enhanced $\ga\ga$ rate in comparison to the SM prediction. Besides the
presence of light scalar taus, in
particular a suppression of the $b\bar b$ decay mode results in an
enhanced $\ga\ga$ rate. This suppression can either be caused by
Higgs-boson propagator corrections entering the effective mixing angle,
or by the so-called $\De_b$ corrections. 

Within the NMSSM the above mentioned mechanisms can also naturally be
realised, and we focused on {\em additional} mechanisms that are 
{\em genuine} for the NMSSM. We found that in particular the
doublet-singlet mixing can result in a substantial suppression of the 
$b \bar b$ mode, resulting again in the desired enhancement in the
$\ga\ga$ rate with respect to the SM prediction.

Finally we have briefly analysed the decays of $\cp$-odd Higgs bosons to
two photons. While an enhancement with respect to the corresponding rate
in the MSSM is possible in the NMSSM, a signal such as reported by ATLAS and
CMS~\cite{discovery} cannot be accounted for by a $\cp$-odd
Higgs.}


\subsection*{Acknowledgements}

We thank  
C.~Duhr, 
B.~Fuks, 
S.~Liebler
and 
F.~Staub
for helpful discussions. We are also grateful to 
T.~Stefaniak 
and 
O.~Brein 
for discussions  and assistance with {\tt HiggsBounds}.
The work of S.H.\ was partially supported by CICYT (grant FPA
2007--66387 and FPA 2010--22163-C02-01), and by the
Spanish MICINN's Consolider-Ingenio 2010 Program under grant MultiDark
CSD2009-00064.
Work supported in part by the
European Community's Marie-Curie Research Training Network under
contract MRTN-CT-2006-035505 ``Tools and Precision Calculations for
Physics Discoveries at Colliders'' and by the Collaborative Research
Center SFB676 of the DFG, ``Particles, Strings, and the Early
Universe''.




\end{document}